\documentclass[12pt]{iopart}
\usepackage[dvips]{graphicx}
\usepackage{amsfonts}
\usepackage{amssymb}
\usepackage{feynmf}
\usepackage{iopams}
\usepackage{setstack}

\begin{document}\setlength{\unitlength}{1mm}
\begin{fmffile}{fm4}

\title[General Eigenvalue Correlations  for the  Real Ginibre Ensemble]{General Eigenvalue Correlations for the  Real Ginibre Ensemble}

\author{Hans-J\"{u}rgen Sommers, Waldemar Wieczorek}

\address{\it Fachbereich Physik, Universit\"{a}t Duisburg-Essen \\
47048 Duisburg, Germany
} \eads{\mailto{H.J.Sommers@uni-due.de}, \mailto{Waldemar.Wieczorek@uni-due.de}}
\begin{abstract}
We rederive in a simplified version the Lehmann-Sommers eigenvalue
distribution for the Gaussian ensemble of asymmetric real  matrices,
invariant under real orthogonal transformations,  as a basis for a
detailed derivation of a Pfaffian generating functional for
$n$-point densities. This produces a simple free-fermion diagram
expansion for the correlations leading to quaternion determinants in
each order n. All will explicitly be given with the help of a very
simple symplectic kernel for even dimension $N$. The kernel is valid
both for complex and real eigenvalues and describes a deep
connection between both. A slight modification by an artificial
additional Grassmannian solves also the more complicated odd-$N$
case. As illustration we present some numerical results in the space
$\mathbb{C}^n$ of complex eigenvalue $n$-tuples.
\end{abstract}
 \submitto{\JPA}

\section{Introduction}
In a recent short communication \cite{Sommers2007} a simple
derivation of the $n$-point eigenvalue correlations for the  real
Ginibre ensemble was announced, which will be presented here in more
detail and extended to most general cases. Ginibre \cite{Ginibre65},
when he proposed his  three types of Gaussian non-Hermitian matrix
ensembles with  complex, quaternion real and real entries
respectively, was not able to solve the correlations for the real
ensemble invariant under orthogonal transformations. It took quite a
long time until Lehmann and Sommers \cite{Lehmann91} derived the
joint probability density of eigenvalues, which is somewhat
difficult to understand since the eigenvalues can be real or
pairwise complex conjugate. Below we will present a simplified
version of the derivation since we need it for obtaining a
generating functional for the correlations, which we present as
symmetric $n$-point densities $R_n (z_1,z_2,\ldots, z_n)$ in the
space of complex eigenvalue $n$-tuples $(z_1,z_2,\ldots, z_n)$.
These include, as singular contributions, correlations with real
eigenvalues and with complex conjugate pairs.\\

\noindent Edelman \cite{Edelman97} rederived the joint probability
density of eigenvalues some years later and determined the 1-point
density of complex eigenvalues. In a somewhat earlier paper Edelman,
Kostlan and Shub \cite{EKS94} calculated  the density of real
eigenvalues and determined their probability. We will use both
results  here to calculate  a very simple skew-symmetric kernel
${\cal K}_N(z_1,z_2)$ which governs all correlation functions. This
kernel follows by a simple argument from Edelman's complex density,
it helps us to invert a large antisymmetric matrix, the result of
which turns out to be simply tridiagonal. This is of course possible
only for even dimension $N$. In this paper we will also show how to
generalize the result for odd $N$. In more recent work Akemann and
Kanzieper
 calculated the smooth complex correlations as Pfaffians \cite{Akemann07}, Sinclair
 derived a generating Pfaffian functional \cite{Sinclair06} and Forrester and Nagao \cite{Forrester07}
 were able to determine the real-real and complex-complex correlations as Pfaffians with the help of
 skew-orthogonal polynomials. Without mentioning the paper \cite{Sommers2007} Borodin and Sinclair
 \cite{Borodin08} generalized
 this to more general and also to crossed correlations. The general solution for odd dimension $N$
 is not contained in these papers.\\

 \noindent Here we will use Grassmannians to derive a simple free fermion zero-dimensional field theory,
which presents the $n$-point density as a diagram expansion, that
turns out to be a quaternion determinant of a $2n \times 2n$
selfdual matrix equivalent to a Pfaffian of a certain skew-symmetric
matrix. The fermionic Wick theorem helps us to work with the
complicated algebra of the Pfaffians. The odd-dimensional case can
be managed by introducing an additional artificial Grassmannian and
again the resulting  large antisymmetric matrix, which we need to
invert for perturbation expansion, turns out to have a simple
tridiagonal inverse. Finally we will present the correlation
functions in a form analytic in the dimension $N$ valid for even and
odd $N$. As illustration we show numerical simulations  for 1-, 2-
and 3- point densities and  numerical presentations of analytical
results for 1- and 2- point correlation
functions.\\

\noindent Let us mention that the real asymmetric Gaussian ensemble
has many applications in physics and social sciences, such as
biological webs \cite{May72}, neural networks \cite{Sompolinsky88}
directed  quantum chaos \cite{Efetov97}, financial markets
\cite{KWAPIEN06} and quantum information theory \cite{Bruzda08}. Our
paper is concerned  with the correlations of the eigenvalues $z_i =
\Lambda_i$ of such real matrices $J_{ij}$ $(1\leqslant i,j \leqslant
N)$ which fulfill the characteristic equation $\det (J_{ij} - z) =
0$ and therefore are real or pairwise complex conjugate.

\section{Joint density of eigenvalues}

We recall the derivation of the joint density of eigenvalues  for
the real Ginibre ensemble, which we simplify in the following. The
normalized measure for the $N$-dimensional real asymmetric random
matrix $J_{ij}$ is given by:
\begin{equation}
\label{1} d\mu (J) = \prod_{i,j}^{1 \ldots N} \left(
\frac{dJ_{ij}}{\sqrt{2 \pi}} \right) \exp \left( - \frac{1}{2}
\sum_{i,j}^{1 \ldots N} J_{ij}^{2} \right)
\end{equation}
For the joint density of eigenvalues $\Lambda_i$ we first consider
the case of dimension $N = 2$, because this  already shows the
essential features.
\subsection{Dimension $N = 2$}
\label{2a} The matrix $ J  = {A\; B \choose C \; D}  = O
\widetilde{J} O^{-1}$ may be brought by an orthogonal transformation
$O$ to the form:
\begin{eqnarray}
  \nonumber \fl  J = \left(\begin{array}{cc}
          \cos \phi & \sin \phi \\
          - \sin \phi & \cos \phi \\
        \end{array} \right)
        \left(\begin{array}{cc}
          \lambda_1 & \delta \\
          -\delta  & \lambda_2 \\
        \end{array}\right)
        \left(\begin{array}{cc}
          \cos \phi & - \sin \phi \\
           \sin \phi & \cos \phi \\
        \end{array}\right) \\
         =  \left( \begin{array}{cc}
          (\lambda_1 + \lambda_2)/2 & \delta \\
          -\delta  & (\lambda_1 + \lambda_2)/2 \\
        \end{array} \right) + \frac{\lambda_1 - \lambda_2}{2} \left(\begin{array}{cc}
          \cos 2\phi & -\sin 2 \phi \\
          - \sin 2 \phi & -\cos 2 \phi \\
        \end{array}\right)  .
\end{eqnarray}
Here the range of parameters is $\lambda_1 > \lambda_2$,  $0
\leqslant \phi \leqslant\pi$, $-\infty < \delta < \infty$.
$\lambda_1, \lambda_2$ are the eigenvalues of the symmetric part of
$J$, $\phi$ is the angle of rotation which diagonalizes the
symmetric part of $J$ and $\delta$ determines the skew-symmetric
part of $J$. Note that ${ 0 \; \; \delta \choose  - \delta \; 0 }$
commutes with ${\cos \phi \; \sin \phi \choose
          - \sin \phi \; \; \; \cos \phi }$ and  thus cannot be brought to diagonal form
by such a transformation. With the skew-symmetric matrix $O^{-1} dO
=  { \; 0 \; d\phi \choose  - d \phi \;  0}$ we obtain
\begin{equation}\label{3}
    dJ = O \bigl( O^{-1} dO \widetilde{J} - \widetilde{J} O^{-1} dO + d \widetilde{J}\bigr) O^{-1} \, .
\end{equation}
For the Jacobian we may drop the rotation of the increment(\ref{3})
and from
\begin{equation}\label{4}
O^{-1} dO \widetilde{J} - \widetilde{J} O^{-1} d O + d \widetilde{J}
= \left(\begin{array}{cc}
          d \lambda_1 & d \delta + d \phi (\lambda_2 - \lambda_1) \\
          -  d \delta + d \phi (\lambda_2 - \lambda_1) & d \lambda_2 \\
        \end{array} \right)
\end{equation}
we find
\begin{equation}\label{5}
    \frac{\partial (A,B,C,D)}{\partial(\phi, \delta,\lambda_1, \lambda_2)} = \left|
    \begin{array}{ccccc}
    0& \lambda_2 - \lambda_1 & \lambda_2 - \lambda_1  & 0\\
    0 & 1 & -1& 0\\
    1 & 0 & 0 & 0\\
    0 & 0 & 0 & 1
    \end{array}  \right| = 2 (\lambda_1 - \lambda_2) \, .
\end{equation}
Thus the measure in the new parametrization is given by
\begin{equation}\label{6}
    d \mu (J) = \frac{1}{(2 \pi)^2} d\phi\, d \delta \, d \lambda_1 \, d \lambda_2\,  2 (\lambda_1 - \lambda_2)
    {\rm e}^{- (\lambda_1^2 + \lambda_2^2 + 2 \delta^2)/2}
\end{equation}
which is positive for $ \lambda_1 > \lambda_2$, $0 \leqslant \phi
\leqslant\pi$, $-\infty < \delta < \infty$.\\
A check of normalization shows that $\int d \mu (J) = 1$.\\

\noindent Let us now go to the eigenvalues $\Lambda_{\pm}$ of $J$
which are obtained from
\begin{equation}\label{7}
    \left| \begin{array}{cc} \lambda_1 - \Lambda& \delta\\
    - \delta & \lambda_2 - \Lambda
    \end{array} \right| = \Lambda^2 - (\lambda_1 + \lambda_2) \Lambda + \lambda_1 \lambda_2 + \delta^2 = 0\, .
\end{equation}
Therefore
\begin{equation}\label{8}
    \Lambda_{\pm} = \frac{\lambda_1 + \lambda_2}{2} \pm \sqrt{\Bigl( \frac{\lambda_1 - \lambda_2}{2}\Bigr)^2 - \delta^2} \, .
\end{equation}
On the other hand we can also write
\begin{equation}\label{9}
    \lambda_{1,2} = \frac{\Lambda_+ + \Lambda_-}{2} \pm \sqrt{\Bigl( \frac{\Lambda_+ - \Lambda_-}{2}\Bigr)^2 + \delta^2}
\end{equation}
with $\lambda_1 > \lambda_2$. Since $\lambda_1,  \lambda_2$ are real
we have
\begin{equation}\label{10}
    \delta^2 \geqslant - \Bigl( \frac{\Lambda_+ - \Lambda_-}{2}\Bigr)^2
\end{equation}
which is only relevant if $\Lambda_{+}$ and $\Lambda_{-}$ are
complex conjugate of each other. $\Lambda_{\pm}$ are either both
real for $(\frac{\lambda_1 - \lambda_2}{2})^2 \geqslant \delta^2$ or
complex for $ \delta^2 > (\frac{\lambda_1 - \lambda_2}{2})^2 $. In
the last case we choose
\begin{equation}\label{11}
    \Lambda_{\pm} = \frac{\lambda_1 + \lambda_2}{2} \pm i \sqrt{\delta^2 - \Bigl( \frac{\lambda_1 - \lambda_2}{2}\Bigr)^2}
\end{equation}
such that $\mbox{ Im} \Lambda_+ > 0$. Now we want to integrate out
$\phi$ and $\delta$ for fixed $\Lambda_+, \Lambda_-$ and find first
\begin{equation}\label{12}
    \frac{\partial (\Lambda_+ , \Lambda_-)}{\partial ( \lambda_1, \lambda_2)} = \frac{\lambda_1 - \lambda_2}{\Lambda_+ -
    \Lambda_-}
\end{equation}
which means

\begin{equation}\label{13}
    (\lambda_1 - \lambda_2) d \lambda_1 \, d \lambda_2 = (\Lambda_+ - \Lambda_-) d\Lambda_+ \,d \Lambda_-
\end{equation}
valid in a sense of an alternating product of differentials, also in
the case if $\Lambda_+$, $\Lambda_-$ are complex. Integrating out
$\phi$ and $\delta$ we obtain for the measure of eigenvalues
\begin{equation}\label{14}
 \fl   d \mu(\Lambda_+ ,  \Lambda_-) = \frac{1}{(2 \pi)^2} \int_0^\pi d\phi \int_{\cal R}
    d \delta \cdot d\Lambda_+ \, d\Lambda_- \; 2 (\Lambda_+ - \Lambda_-) {\rm e}^{- (\Lambda_+^2 + \Lambda_-^2 + 4 \delta^2)/2}
\end{equation}
where $\delta$ is integrated over the region $\cal R$ given by eq.
(\ref{10}). If $\Lambda_\pm$ are real this yields simply
\begin{equation}\label{15}
d \mu(\Lambda_+ ,  \Lambda_-) = \frac{1}{2\sqrt{ 2 \pi}}  d\Lambda_+
\, d\Lambda_- \;  (\Lambda_+ - \Lambda_-)
 {\rm e}^{- (\Lambda_+^2 + \Lambda_-^2)/2}
\end{equation}
and if $\Lambda_\pm$ are complex
\begin{equation}\label{16}
\fl d \mu(\Lambda_+ ,  \Lambda_-) = \frac{1}{2\sqrt{ 2 \pi}}
d\Lambda_+ \, d\Lambda_- \;  (\Lambda_+ - \Lambda_-)
 {\rm e}^{- (\Lambda_+^2 + \Lambda_-^2)/2} \mbox{ erfc} ( | \mbox{ Im} \Lambda_+ | \sqrt{2})
\end{equation}
with $\mbox{ erfc}(z) = \frac{2}{\sqrt{\pi}} \int_z^\infty dx \,
{\rm e}^{-x^2}$. Formula  (\ref{16}) is in this form also valid for
the real case. This measure is positive, since in the real case we
assume $\Lambda_+ > \Lambda_-$ and in the complex case  $\Lambda_\pm
= x \pm iy$ we have $d\Lambda_+ \, d\Lambda_- = (dx + idy)(dx - i
dy) = - 2 i\, dx dy$ and $y>0$.\\

\noindent We can also check the normalization. For the real
eigenvalues we obtain
\begin{equation}\label{17}
    \int_{\Lambda_\pm \mbox{ real}} d\mu(\Lambda_+, \Lambda_-) = \frac{1}{\sqrt{2}} \, .
\end{equation}
This is the probability that both eigenvalues are real. For the
complex part we obtain correctly
\begin{equation}\label{18}
    \int_{\Lambda_\pm \mbox{ complex}} d\mu(\Lambda_+, \Lambda_-) = 1 - \frac{1}{\sqrt{2}}
\end{equation}
such that total probability is 1.
\subsection{General derivation}
We start again from eq. (\ref{1}). The dimension $N$ is even or odd,
we will see that the odd case is more complicated. Again we will
bring $J$ by an orthogonal transformation $O$ to a kind of lower
triangular form
\begin{equation}\label{19}
    J = O \; \widetilde{J} \; O^{-1} = O ( \Delta + \Lambda ) \;  O^{-1}
\end{equation}
with in the even case $\frac{1}{2} N\;$  $2 \times 2$ blocks
$\Lambda$ in the diagonal  and in the odd case one $1\times 1$ block
together with $\frac{1}{2} (N-1)\;$ $2 \times 2$ blocks in the
diagonal. Then we obtain again eq. (\ref{3}) for the increment $dJ$.
Since $O^{-1} dO$ is skew-symmetric we can consider $d\widetilde{J}$
and $(O^{-1} dO \widetilde{J} - \widetilde{J} O^{-1}
dO)_{\mbox{\tiny above}}$ (i.e. where $\widetilde{J}$ is zero) as
independent variables. The latter may be reduced to $(O^{-1} dO
\Lambda - \Lambda O^{-1} dO)_{\mbox{ \tiny above}}$ with Jacobian
$1$ due to the triangular structure of $\Delta$, and these again to
$(O^{-1} dO)_{\mbox{ \tiny  above}}$ with Jacobian $\prod_{i<j}^{'}
(\Lambda_i - \Lambda_j)$ where $\Lambda_i$ are the eigenvalues of
$\Lambda$. Here $\prod^{'}$ means that  the  blocks in the diagonal
are excluded: $\prod_{i<j}^{'} = \prod_{\mbox{\tiny above}}$. If the
blocks $\Lambda$ are themselves diagonal this is easy to see. But it
is then even generally true by diagonalization of the blocks
$\Lambda$
in the generic case.\\

\noindent Thus  we integrate first over $O(N)/O(2)^{N/2}$ in the
even-$N$ case and over $O(N)/O(2)^{(N-1)/2} \times O(1)$ in the
odd-$N$  case, that is over the orthogonal group modulo rotations
which leave the diagonal block structure invariant. Then we
integrate over $\Delta$. For the remaining  blocks in the diagonal
we can use our 2-dimensonal result from Section \ref{2a}. Finally
we arrive at the result derived by Lehmann and Sommers
\cite{Lehmann91} and rediscovered by Edelman \cite{Edelman97}
\begin{eqnarray}\label{20}
\nonumber
 \fl   d\mu(\Lambda_1,\Lambda_2, \ldots, \Lambda_N) =& K_N \cdot  d\Lambda_1 \ldots d\Lambda_N  \cdot \\
    & \cdot \prod_{i<j} (\Lambda_i - \Lambda_j) \cdot
    {\rm e}^{-\sum_i \Lambda_i^2/2} \cdot \Bigl( \prod_i \mbox{ erfc} ( |\mbox{ Im} \Lambda_i| \sqrt{2}) \Bigr)^{1/2}
\end{eqnarray}
with
\begin{equation}\label{20a}
K_N = VO(N) \cdot 2^{-N} (2 \pi)^{-N(N+1)/4} \, .
\end{equation}
 The constant $K_N$
comes from $VO(N)/(4 \pi)^{N/2}$ in the even case and $VO(N)/(4
\pi)^{(N-1)/2} \cdot 2$ in the odd case and integration over
$\Delta$. In both cases the result is the same.
\begin{equation} \label{20b}
VO(N) = \prod_{d=1}^N \frac{2 \pi^{d/2}}{\Gamma(d/2)} = \int \Bigl|
\prod_{i<j} (O^{-1} dO)_{ij} \Bigr|
\end{equation}
 is the volume
of the $N$-dimensional orthogonal group.\\

\noindent Here we have to assure that the transformation is unique.
Thus the eigenvalues $\Lambda_i$ have to be different and arranged
in a special order. If all eigenvalues are real we assume $\Lambda_1
> \Lambda_2 > \Lambda_3 \ldots$. If two eigenvalues are complex we
assume $\Lambda_1 = \overline{\Lambda_2}$ and $\mbox{ Im} \Lambda_1
= - \mbox{ Im} \Lambda_2 > 0$ and  $\Lambda_3 > \Lambda_4
> \Lambda_5 \ldots$. If 4 eigenvalues are complex we assume
$\Lambda_1 = \overline{\Lambda_2}$, $\Lambda_3 =
\overline{\Lambda_4}$, $\mbox{ Re} \Lambda_1 > \mbox{ Re}
\Lambda_3$, $\mbox{ Im} \Lambda_1, \mbox{ Im} \Lambda_3 > 0$ and
$\Lambda_5 > \Lambda_6 \ldots$ And so on. The measure is written in
such a way that it is positive even in the complex case for which
e.g. $\Lambda_1 = \overline{\Lambda_2} = x_1
+ i y_1$.\\

\noindent Now we want to determine the correlation functions. To
this end we go over to two-dimensional variables
\begin{equation}\label{21}
    \Lambda\rightarrow z = x + i y \mbox{ with } d^2 z = dx \, dy\, .
\end{equation}
Then we ask for the symmetrized probability $P(z_1,z_2, \ldots, z_N)
d^2z_1 \ldots d^2z_N$ that the complex eigenvalue tuple
$(z_1,z_2,\ldots,z_N)$ appears in the volume element $d^2z_1 \ldots
d^2z_N$, where now we drop the restrictions for $\Lambda_i = z_i$.
\section{Generating functional}
\subsection{Definition}
We are interested in the $n$-point  densities (or correlation
functions)
\begin{equation}\label{22}
    R_n(z_1,z_2,\ldots,z_n) = \left. \frac{\delta}{\delta f(z_1)} \ldots \frac{\delta}{\delta f(z_n)} Z[f] \right|_{f\equiv1}
\end{equation}
with
\begin{equation}
\label{23} Z[f] = \int d^2z_1 \ldots d^2z_N \; P(z_1,z_2,\ldots,z_N)
f(z_1) \ldots f(z_N)\; .
\end{equation}
Recall that the variables $\Lambda_k = z_k = x_k + i y_k$ are
considered here as two dimensional vectors. We obtain $Z[f]$ by
integrating over our joint density (\ref{20}). Here we have to
distinguish between the different cases: all $z_k$ real, 2
eigenvalues complex, 4 eigenvalues complex and so on:
\begin{eqnarray}\label{24}
 \nonumber   \fl  Z[f]  =  K_N \int d^2z_1 \ldots d^2z_N f(z_1) \ldots
 f(z_N) \prod_{i<j} (z_i - z_j) \prod_{k}
  {\rm e}^{-z_k^2/2} (\mbox{ erfc} ( |\mbox{ Im} z_k |\sqrt{2} ))^{1/2}\\
  \nonumber  \cdot \left\{ \delta(y_1) \delta(y_2) \ldots  \delta(y_N) \;   \Theta(x_1 > x_2 > \ldots > x_N) \right.\\
  \nonumber  + (-2i) \Theta(y_1)  \delta(y_1 + y_2)  \delta(x_1 - x_2) \delta(y_3) \ldots  \delta(y_N)
  \Theta(x_3 > \ldots > x_N)\\
  \nonumber  + (-2i)^2 \Theta(y_1)  \Theta(y_3) \delta^2(z_1 - \overline{z_2}) \delta^2 (z_3 - \overline{z_4}) \Theta (x_1 - x_3)\\
   \left. \cdot \delta(y_5)  \delta(y_6) \ldots  \delta(y_N)  \Theta(x_5 > x_6 > \ldots > x_N) + \ldots \right\}
\end{eqnarray}
 By integrating over
with the symmetric function $f(z_1) \ldots f(z_N)$ the integrand
will automatically be symmetrized. We used the notation
$\delta^2(z_1 - \overline{z_2}) =  \delta(y_1 + y_2)  \delta(x_1 -
x_2)$. $\Theta(x_1 > x_2 > \ldots > x_N)$ stands for the restriction
$x_1 > x_2 > \ldots > x_N$.
\subsection{Vandermonde determinant}
\label{3b} Let us write the Vandermonde determinant as
\begin{equation}\label{25}
   \fl  \prod_{i<j} (z_i - z_j) = (-1)^{N(N-1)/2} \prod_{i > j} (z_i - z_j) = (-1)^{\frac{N(N-1)}{2}} \det [z_1^{k-1},z_2^{k-1}
    ,\ldots, z_n^{k-1} ]
\end{equation}
with $k = 1, 2, \ldots, N$. The determinant can be written with
Grassmannian variables $\eta^{*}_k, \eta_l$ and Berezin integration
\begin{equation}\label{26}
 \prod_{i>j} (z_i - z_j) = \int d\eta_1^* d\eta_1 \dots d\eta_N^* d\eta_N \exp \Bigl( -\sum_{kl}
 \eta_k^* z_l^{k-1} \eta_l \Bigr)
\end{equation}
Integrating out one set of Grassmannians $(\eta_l)$ we obtain:
\begin{equation}\label{27}
\fl  \prod_{i<j} (z_i - z_j) = \int d\eta_1^* d\eta_2^* \ldots
d\eta_N^* \Bigl(\sum_k \eta_k^* z_1^{k-1}\Bigr) \Bigl(\sum_k
\eta_k^* z_2^{k-1}\Bigr)
 \dots \Bigl(\sum_k \eta_k^* z_N^{k-1}\Bigr) \, .
\end{equation}
Note that the integrand factorizes in a product of identical
functions of different arguments $(z_i)$.
\subsection{Real case}
In the case that all eigenvalues are real we have to integrate a
function $\widetilde{f}(x_1) \ldots \widetilde{f} (x_N)$, where
$\widetilde{f}(x)$ is actually Grassmannian from section \ref{3b}.
with the restriction
\begin{equation}\label{28}
    \Theta (x_1 - x_2) \Theta(x_2 - x_3) \ldots \Theta(x_{N-2} - x_{N-1}) \Theta(x_{N-1}-x_N)
\end{equation}
where $\Theta(x)$ is the Heaviside step function = 1 for $x>0$ and 0
for $x<0$.\\
Using Mehta's method of alternating variables we integrate first
over $x_N, x_{N-2}, x_{N-4}$  and so on and obtain an integral
\begin{equation}\label{29}
    I  = \int \ldots  \widetilde{f}(x_{N-3})  \int_{x_{N-1}}^{x_{N-3}} dx_{N-2} \widetilde{f}(x_{N-2}) \widetilde{f}(x_{N-1}) \int_{-\infty}^{x_{N-1}}
    d x_N \widetilde{f} (x_N) \, .
\end{equation}
Now we use that $\widetilde{f}(x)$ is Grassmannian and therefore
also $\int_{-\infty}^x dx_N \widetilde{f} (x_N)$ for which the
square vanishes. Thus we may replace the above result by
\begin{equation}\label{30}
    I = \int \ldots  \widetilde{f}(x_{N-3}) \int_{-\infty}^{x_{N-3}} dx_{N-2} \widetilde{f}(x_{N-2}) \widetilde{f}(x_{N-1}) \int_{-\infty}^{x_{N-1}}
   dx_N \widetilde{f} (x_N) \, .
\end{equation}
There remains the restriction $x_1
> x_3 > \ldots > x_{N-1}$ for even $N$, dropping it we obtain
\begin{equation}\label{31}
    I = \frac{1}{(N/2)!} \Bigl[ \int_{-\infty}^{+\infty} dx_1 \widetilde{f}(x_1) \int_{-\infty}^{x_1} dx_2
    \widetilde{f} (x_2) \Bigr]^{N/2} \, .
\end{equation}
And if $N$ is odd we have
\begin{equation}\label{32}
    I = \int_{-\infty}^{+\infty} dx \widetilde{f} (x) \frac{1}{((N-1)/2)!}
    \Bigl[ \int_{-\infty}^{+\infty} dx_1 \widetilde{f}(x_1) \int_{-\infty}^{x_1} dx_2
    \widetilde{f} (x_2) \Bigr]^{(N-1)/2} \, .
\end{equation}
\subsection{General case}
Now we first  assume that $N$ is even and all eigenvalues are
pairwise complex conjugate. We have the restriction (besides $z_1 =
\overline{z_2}, \; z_3 = \overline{z_4}$ etc; $y_1, y_3, \ldots ,
y_{N-1} > 0$)
\begin{equation}\label{33}
     \Theta(x_1 - x_3) \Theta(x_5 - x_7) \dots \Theta(x_{N-3} - x_{N-1}) \, .
\end{equation}
Dropping the restriction we obtain for the integral over
$\widetilde{f}(z_1) \ldots \widetilde{f}(z_N)$
\begin{equation}\label{34}
    I = \frac{1}{(N/2)!} \Bigl[ -2i \int d^2z \Theta(y)  \widetilde{f}(z) \widetilde{f}(\overline{z}) \Bigr]^{N/2} \, .
    \end{equation}
Now it is easy to sum over all mixed cases. For even $N$ we obtain
\begin{eqnarray}\label{35}
 \nonumber I &=   \frac{1}{(N/2)!} \sum_{M=0}^{N/2} \left\{ \Bigl( \begin{array}{c} N/2\\
M
\end{array}\Bigr) \Bigl[ -2i \int d^2z \Theta(y) \widetilde{f}(z)
\widetilde{f}(\overline{z}) \Bigr]^{M} \cdot \right. \\
\nonumber
& \parbox{3cm}{\hspace*{3cm}} \left.  \cdot \Bigl[
\int_{-\infty}^{+\infty} dx_1  \widetilde{f}(x_1)
\int_{-\infty}^{x_1} dx_2 \widetilde{f} (x_2) \Bigr]^{\frac{N}{2} - M} \right\}\\
 \nonumber &= \frac{1}{(N/2)!} \Bigl[ -2i \int d^2z \Theta(y)
\widetilde{f}(z) \widetilde{f}(\overline{z})  + \\
& \parbox{3cm}{\hspace*{3cm}} + \int_{-\infty}^{+\infty} dx_1 \widetilde{f}(x_1)
\int_{-\infty}^{x_1} dx_2 \widetilde{f} (x_2) \Bigr]^{N/2}
\end{eqnarray}
and similarly in the odd case we obtain
\begin{eqnarray}  \nonumber
 I =& \int_{- \infty}^{+ \infty}  dx \widetilde{f}(x)
\frac{1}{(\frac{N-1}{2})!} \Bigl[ -2i \int d^2z \Theta(y)
\widetilde{f}(z) \widetilde{f}(\overline{z}) + \\
\label{36}
& \qquad + \int_{-\infty}^{+\infty} dx_1 \widetilde{f}(x_1)
\int_{-\infty}^{x_1} dx_2 \widetilde{f} (x_2) \Bigr]^{(N-1)/2}
\end{eqnarray}
\subsection{$Z[f]$ as a Pfaffian}
Now we may write the generating functional as integral over
Grassmannians (in the even-$N$ case)
\begin{eqnarray}\label{37}
  \nonumber  Z[f] &=& K_N \int d\eta_1^* d\eta_2^* \ldots d\eta_N^* \; \frac{1}{(N/2)!} \left[ - \frac{1}{2} \sum_{kl}^{1\ldots N}
    \eta_k^* \widetilde{A}_{kl} \eta_l^* \right]^{N/2} \\
     &=& K_N \int d\eta_1^* d\eta_2^* \ldots d\eta_N^* \; \exp \Bigl(- \frac{1}{2} \sum_{kl}
    \eta_k^* \widetilde{A}_{kl} \eta_l^* \Bigr)
\end{eqnarray}
with the skew-symmetric matrix
\begin{equation}\label{38}
\widetilde{A}_{kl} = \int d^2 z_1 d^2 z_2 f(z_1) f(z_2) {\cal F}
(z_1, z_2) z_1^{k-1} z_2^{l-1}
\end{equation}
and the skew-symmetric measure
\begin{eqnarray}\label{39}
\nonumber   \fl  {\cal F} (z_1, z_2)  =& {\rm e}^{-(z_1^2 +
z_2^2)/2} \left[ 2 i \delta^2 (z_1 - \bar{z}_2)
    \{ \Theta(y_1) \mbox{ erfc} (y_1\sqrt{2}) \right. \\
&- \left. \Theta(y_2) \mbox{ erfc} (y_2 \sqrt{2}) \} + \delta(y_1)
\delta(y_2) \left(\Theta (x_2 - x_1) - \Theta (x_1 - x_2) \right)
\right] \; .
\end{eqnarray}
We have antisymmetrized because $\eta_k^*$ are Grassmannians. Such
an expression for $Z[f]$ is called a Pfaffian
\begin{equation}\label{40}
    Z[f] = K_N \mbox{ Pfaff} \bigl( \widetilde{A} \bigr) = K_N \sqrt{\det(\widetilde{A})} \, .
\end{equation}
The Pfaffian is an analytic square root of the determinant $\det
(\widetilde{A})$. It is only defined for an antisymmetric matrix. In
this case we know that for positive $f(z)$ $\mbox{ Pfaff}
(\widetilde{A})$ is positive, such that the square root of the
positive determinant $\det (\widetilde{A})$ is also positive. We
actually need $f(z)$ only in an infinitesimal region near
$f(z)\equiv 1$.\\

\noindent We may immediately consider odd $N$. In that case
$\widetilde{A}_{kl}$ has always a zero eigenvalue and cannot be
inverted (what we will need for perturbation expansions). Therefore
we  increase artificially the number of Grassmannians by $1$
\begin{eqnarray}\label{41}
 \fl   \nonumber  Z[f] = K_N \int d\eta_1^*  \ldots d\eta_{N+1}^* \;\eta_{N+1}^* \int dx f(x) {\rm e}^{-x^2/2}
    \sum_{k=1}^{ N}\eta_k^* x^{k-1} {\rm e}^{-\frac{1}{2}
      \sum_{kl}^{1\ldots N}
    \eta_k^* \widetilde{A}_{kl} \eta_l^{*}} \\
    = K_N \int d\eta_1^*  \ldots d\eta_{N+1}^* \; \exp \Bigl( - \frac{1}{2} \sum_{nm}
    \eta_n^* \widetilde{B}_{nm} \eta_m^* \Bigr)
\end{eqnarray}
Now $n,m = 1,2, \ldots, N+1$ and
\begin{equation}\label{42}
    \widetilde{B}_{n,m} = \left[ \begin{array}{cc} \widetilde{A}_{kl} & \widetilde{C}_k\\
    -\widetilde{C}_l & 0
    \end{array} \right]
\end{equation}
with
\begin{equation}\label{43}
\widetilde{C}_k = \int dx f(x) {\rm e}^{-x^2/2} x^{k-1} \, .
\end{equation}
Thus we obtain in the odd-$N$ case
\begin{equation}\label{44}
    Z[f] = K_N \mbox{ Pfaff} (\widetilde{B}_{nm})
\end{equation}
The correlation functions can be found by multiple derivating of
$Z[f]$ w.r.t. $f(z)$ at $f(z) \equiv 1$.
\section{1-point density}
To calculate the 1-point density we use
\begin{equation}\label{45}
    R_1(z) = \left. \frac{\delta Z[f]}{\delta f(z)} \right|_{f \equiv 1}
    = \left. \frac{\delta \ln Z[f]}{\delta f(z)} \right|_{f \equiv 1} \, .
\end{equation}
\subsection{Even $N$, complex eigenvalues}
In the even-$N$ case we have
\begin{equation}\label{46}
    R_1(z) =  \frac{1}{2} \frac{\delta \ln \det \widetilde{A}}{\delta f(z)} = \frac{1}{2} \mbox{ Tr }
    \frac{1}{\widetilde{A}} \left. \frac{\delta  \widetilde{A}}{\delta f(z)} \right|_{f \equiv 1}
\end{equation}
where $\widetilde{A}_{kl}$ is given by eq. (\ref{38}). Let us call
$A_{kl} = \widetilde{A}_{kl}|_{f \equiv 1}$ and introduce  the
kernel
\begin{equation}\label{47}
    {\cal K}_N(z_2,z_1) = \sum_{k,l}^{1 \ldots N} A^{-1}_{kl} z_2^{k-1} z_1^{l-1} \, .
\end{equation}
Then we obtain
\begin{equation}\label{48}
    R_1(z_1) = \int d^2z_2 \, {\cal F} (z_1,z_2) {\cal K}_N(z_2,z_1)
\end{equation}
with ${\cal F} (z_1,z_2)$ given by eq. (\ref{39}). If we insert eq.
(\ref{39}) we obtain
\begin{equation}\label{49}
    R_1(z_1) = R_1^C(z_1) + \delta(y_1) R_1^R(x_1)
\end{equation}
with
\begin{equation}\label{50}
    R_1^C(z_1) = 2 i \mbox{ sgn } (y_1) \mbox{ erfc} ( |y_1 |\sqrt{2}) {\rm e}^{-  x_1^2 + y_1^2}
    {\cal K}_N(\overline{z_1}, z_1)
\end{equation}
and
\begin{equation}\label{51}
    R_1^R (x_1) = \int_{-\infty}^{+\infty} dx_2 \mbox{ sgn } (x_2 - x_1) {\rm e}^{- (x_1^2 + x_2^2)/2} {\cal K}_N(x_2,x_1) \, ,
\end{equation}
which  is a smooth  part $R_1^C (z_1)$ in the complex plane and a
part $R_1^R(x_1)$ concentrated on the real axis.\\

\noindent  If we compare $R_1^C (z_1)$ with Edelman's expression
\cite{Edelman97}  for the complex 1-point density we find for
${\cal K}_N (z_2,z_1)$ using that $\overline{z_1}$ and $z_1$ are
independent variables
\begin{equation}\label{54a}
    {\cal K}_N (z_2,z_1) = \frac{z_2 - z_1}{2 \sqrt{2 \pi}} \sum_{n=0}^{N-2} \frac{(z_1 z_2)^n}{n!} = \sum_{k,l}^{1 \ldots N}
    A_{kl}^{-1} z_2^{k-1} z_1^{l-1}\, .
\end{equation}
Thus surprisingly, the skew symmetric matrix $A^{-1}_{kl}$ has a
very simple tridiagonal structure
\begin{equation}\label{52}
    A^{-1}_{kl} = \frac{1}{2\sqrt{2\pi}} \left[ \begin{array}{ccccc}
    0& - \frac{1}{0!} & 0 & \ldots & 0\\
    \frac{1}{0!} & 0 & -\frac{1}{1!} & \ddots & \vdots\\
    0 & \frac{1}{1!}& \ddots & \ddots &0\\
    \vdots& \ddots& \ddots & \ddots & - \frac{1}{(N-2)!}\\
    0& \ldots & 0 & \frac{1}{(N-2)!}& 0
    \end{array} \right]
\end{equation}
which leads to Edelman's expression
\begin{equation}\label{53}
    R_1^C (z)  = \frac{2 |y|}{\sqrt{2 \pi}} \mbox{ erfc } ( |y| \sqrt{2}) {\rm e}^{-x^2 + y^2}
    \sum_{n=0}^{N-2} \frac{|z|^{2 n}}{n!} \, .
\end{equation}
For $N=2$ it agrees with our 2-dimensional expression eq.
(\ref{16}). Note that $R_1 (z)$ as a density is normalized to $N$:
$\int d^2z R_1(z) = N$. Similarly $\int d^2 z_1 d^2z_2 R_2(z_1,z_2)
= N (N-1)$
and so on.\\
Using the formula
\begin{equation}\label{54}
    {\rm e}^{-v} \sum_{n=0}^{N} \frac{v^n}{n!} = \int_v^{\infty} du\; {\rm e}^{-u} \frac{u^{N}}{N!}
\end{equation}
we can also write
\begin{equation}\label{55}
    R_1^C (z) = \frac{2 |y|}{\sqrt{2 \pi}} \mbox{ erfc } (\sqrt{2} |y|) {\rm e}^{2 y^2} \int_{|z|^2}^{\infty}
    du \, {\rm e}^{-u} \frac{u^{N-2}}{\Gamma(N-1)}
\end{equation}
analytic in $N$. From (\ref{52}) it is easy to check the
normalization (eqs. (\ref{20a},\ref{20b})) using the duplication
formula for the Gamma function. Note that $\mbox{Pfaff } (A_{kl})$
is positive.
\subsection{Even $N$, real eigenvalues}
\label{4b} Since the same analytic kernel appears also in the
density of real eigenvalues eq. (\ref{51}) we have using eq.
(\ref{54})
\begin{equation}\label{56}
 \fl   R_1^R (x_1) = \frac{1}{2 \sqrt{2 \pi}} \int dx_2 \mbox{ sgn } (x_2 - x_1) {\rm e}^{- \frac{(x_1 - x_2)^2}{2}}
    (x_2 - x_1) \int_{x_1 x_2}^{\infty} du {\rm e}^{-u} \frac{u^{N-2}}{(N-2)!}\, .
\end{equation}
Integration by parts yields
\begin{eqnarray}\label{57}
   \nonumber R_1^R (x_1) &= \frac{1}{\sqrt{2\pi}}
\int_{|x_1|^2}^{\infty} du \frac{{\rm e}^{-u} u^{N-2}}{(N-2)!} -\\
\nonumber
&\qquad    - \frac{1}{2 \sqrt{2\pi}} \int_{-\infty}^{+\infty} dx_2 \mbox{ sgn } (x_2 - x_1)  \frac{x_1 (x_1 x_2)^{N-2}}{(N-2)!}
    {\rm e}^{- (x_1^2 + x_2^2)/2}\\
    &= \widetilde{R}_1^R (x_1) - D_N x_1^{N-1} {\rm e}^{-x_1^2/2} \Theta (N \mbox{ odd})
\end{eqnarray}
with
\begin{eqnarray}\nonumber
 \widetilde{R}_1^R (x_1) =& \frac{1}{\sqrt{2 \pi}}
\int_{|x_1|^2}^{\infty} du {\rm e}^{-u} \frac{u^{N-2}}{\Gamma(N-1)}
+  \\
\label{58}
& \quad  +\frac{1}{\sqrt{2\pi}} \int_0^{|x_1|} dx\,   {\rm e}^{-x^2/2}
\frac{x^{N-2} }{\Gamma(N-1)} |x_1|^{N-1} {\rm e}^{-x_1^2/2}
\end{eqnarray}
and
\begin{equation}\label{59}
    D_N = \int_0^\infty \frac{dx {\rm e}^{-x^2/2} x^{N-2}}{\sqrt{2 \pi} \Gamma(N-1)} = \frac{1}{2^{N/2} \Gamma(N/2)}\, .
\end{equation}
The second term in the last line of eq. (\ref{57}) appears formally
only if $N$ is odd. We see that $\widetilde{R}_1^R (x_1)$ is an
analytic function of $N$ which is manifestly  positive. This result
has been obtained by Edelman, Kostlan and Shub \cite{EKS94} and is
also valid for odd $N$. We have to subtract just the second term in
(\ref{57}) which is proportional to $\Theta(N \mbox{ odd})$ (which
means that this is only 1 if $N$ is odd otherwise it is zero) to get
the correct answer valid for even and odd $N$. We will see the
consequences in the following.
\subsection{Odd $N$}
If $N$ is odd $R_1(z_1)$ is given by
\begin{eqnarray}
  \nonumber R_1(z_1) &=& \left. \frac{1}{2} \mbox{ Tr } \frac{1}{\widetilde{B}}
  \frac{\delta \widetilde{B}}{\delta f(z_1)}  \right|_{f \equiv 1} \\
   \nonumber &=&  \left. \left( \frac{1}{2} \sum_{k,l}^{1\ldots N} B^{-1}_{kl}
   \frac{\delta \widetilde{A}_{lk}}{\delta f(z_1)}
     + \sum_{l=1}^{N} B^{-1}_{N+1,l} \; \frac{\delta \, \widetilde{C_{l}}}{\delta f(z_1)} \right) \right|_{f\equiv 1} \\
   &=& \int d^2 z_2 {\cal F} (z_1,z_2) {\cal K}_N (z_2,z_1) + \sum_{l=1}^{N} B^{-1}_{N+1,l} \delta(y_1)
   {\rm e}^{-x_1^2/2} x_1^{l-1} \, .
\end{eqnarray}
Since the complex-eigenvalue part $R_1^C (z_1)$ is again given by
Edelman's expression (\ref{53}) and the real-eigenvalue  part $R_1^R
(x_1)$ is given by eq. (\ref{58}) we conclude that $B^{-1}_{nm}$ has
again a very simple tridiagonal structure
\begin{equation}\label{61}
    B_{nm}^{-1} = \frac{1}{2 \sqrt{2 \pi}} \left[ \begin{array}{cccccc}
    0&-\frac{1}{0!} & 0 & \ldots & \ldots &0\\
    \frac{1}{0!}& 0 &- \frac{1}{1!}&\ddots&&\vdots\\
    0&\frac{1}{1!}&\ddots&\ddots&\ddots & \vdots\\
    \vdots&\ddots&\ddots&\ddots&\frac{-1}{(N-2)!}&0\\
    \vdots&&\ddots&\frac{1}{(N-2)!}&0&{\small -2 \sqrt{2 \pi}D_N}\\
    0&\cdots&\cdots&0&2 \sqrt{2 \pi} D_N&0
    \end{array}
    \right] \, ,
\end{equation}
only the last row and column make the matrix $B^{-1}$ invertible.
The kernel now is given by
\begin{equation}\label{62}
    {\cal K}_N (z_2,z_1) = \sum_{k,l}^{1\ldots N} B^{-1}_{k, l} z_2^{k-1} z_1^{l-1} = \frac{z_2 - z_1}{2 \sqrt{2 \pi}}
    \sum_{n=0}^{N-2} \frac{(z_2 z_1)^n}{n!}
\end{equation}
while
\begin{equation}\label{63}
    B^{-1}_{N+1,k} = - B^{-1}_{k,N+1} = \delta_{k,N} \frac{1}{2^{N/2} \Gamma{(N/2)}} = \delta_{k,N} D_N \, .
\end{equation}
In a diagrammatic representation, which we will use later, we have
with an obvious notation
\begin{eqnarray}
 \nonumber R_1(z_1) &=& \quad
 \parbox{10mm}{
 \begin{fmfchar*}(10,10)
 \fmfpen{thin}
 \fmfset{wiggly_len}{2mm}
 \fmfbottom{i1} \fmftop{o1}
 \fmflabel{$1$}{o1}
 \fmfdot{o1} \fmfdot{i1}
  \fmf{plain,left,tension=0.5}{i1,o1} \fmf{wiggly,left,tension=0.5}{o1,i1}
 \end{fmfchar*}}
+  \quad
\parbox{25mm}{
 \begin{fmfchar*}(10,10)
 \fmfpen{thin}
 \fmfset{wiggly_len}{2mm}
 \fmfbottom{i1} \fmftop{o1}
 \fmflabel{$1$}{o1}
 \fmfdot{o1} \fmfdot{o1}
  \fmfv{decor.shape=cross,decoration.size=2.5mm}{i1}
  \fmf{plain,left,tension=0.5}{i1,o1} \fmf{wiggly,left,tension=0.5}{o1,i1}
 \end{fmfchar*}}
  \\
 &=& \int_{2} {\cal F}(1,2) {\cal K}_N(2,1) + \delta(y_1)
\frac{x_1^{N-1} {\rm e}^{-x_1^2/2}}{2^{N/2} \Gamma(N/2)} \Theta(N
\mbox{ odd}) \, .
 \end{eqnarray}
 Note that the last term gives exactly the correct density for $N=1$.
\section{Correlation functions}
\subsection{Diagram expansion}
Let us write $f(z) = 1 + u(z)$ and
\begin{equation}\label{65}
    \widetilde{A}_{kl} = A_{kl} + C_{kl} = \int d^2z_1 d^2z_2 \, z_1^{k-1} z_2^{l-1} {\cal F} (z_1,z_2)
    (1 + u(z_1))(1+ u(z_2))
\end{equation}
and try to expand $Z[1 + u]$ in powers of $u(z)$ to get the
$n$-point densities. To this end we first expand $Z[1 + u]$ in
powers of $C_{kl}$ (first the even-$N$ case):
\begin{eqnarray}
 \nonumber 
  Z[1+u] &=& K_N \int d\eta_1^* \ldots d \eta_N^* \exp \Bigl( - \frac{1}{2} \sum_{kl} (A_{kl} + C_{kl})
   \eta_k^* \eta_l^* \Bigr)  \\
   &=& Z_0 \sum_{n=0}^{\infty} \frac{1}{n!} \left< \Bigl( - \frac{1}{2} \sum_{kl} C_{kl} \eta_k^* \eta_l^* \Bigr)^n
   \right>_0
\end{eqnarray}
We use now that there is a fermionic Wick theorem with
\begin{eqnarray}
  \langle \eta_k^* \eta_l^* \rangle_{0} = A_{kl}^{-1}&=& \quad
  \parbox{25mm}{\begin{fmfchar*}(15,10)
 \fmfpen{thin}
 \fmfset{wiggly_len}{2mm}
 \fmfleft{i1} \fmfright{o1}
 \fmflabel{$k$}{i1}
 \fmflabel{$l$}{o1}
 \fmf{plain}{i1,o1}
 \fmfdot{i1}
 \fmfdot{o1}
 \end{fmfchar*}}
 \\
  - C_{kl}&=& \quad
  \parbox{25mm}{\begin{fmfchar*}(15,10)
 \fmfpen{thin}
 \fmfset{wiggly_len}{2mm}
 \fmfleft{i1} \fmfright{o1}
 \fmflabel{$k$}{i1}
 \fmflabel{$l$}{o1}
 \fmf{wiggly}{i1,o1}
 \fmfdot{i1}
 \fmfdot{o1}
 \end{fmfchar*}}
\end{eqnarray}
such that with the linked cluster theorem
\begin{equation}\label{69}
    \ln (Z[1+u]/Z_0) = \quad
    \parbox{10mm}{\begin{fmfchar*}(10,10)
 \fmfpen{thin}
 \fmfset{wiggly_len}{2mm}
 \fmftop{i1} \fmfbottom{o1}
 \fmfdot{i1}
 \fmfdot{o1}
 \fmf{plain,right,tension=0.5}{i1,o1}
 \fmf{wiggly,right,tension=0.5}{o1,i1}
 \end{fmfchar*}}
  \quad + \quad
    \parbox{10mm}{\begin{fmfchar*}(10,10)
 \fmfpen{thin}
 \fmfset{wiggly_len}{2mm}
 \fmfsurround{i1,i2,i3,i4}
 \fmfdot{i1,i2,i3,i4}
 \fmf{plain,right=.5,tension=0.35}{i1,i2}
 \fmf{plain,right=.5,tension=0.25}{i3,i4}
 \fmf{wiggly,right=.5,tension=0.25}{i2,i3}
 \fmf{wiggly,right=.5,tension=0.25}{i4,i1}
 \end{fmfchar*}}
 \quad + \quad
 \parbox{10mm}{\begin{fmfchar*}(10,10)
 \fmfpen{thin}
 \fmfset{wiggly_len}{2mm}
 \fmfsurroundn{i}{6}
 \fmfdotn{i}{6}
 \fmf{plain,right=.33,tension=0.25}{i1,i2}
 \fmf{plain,right=.33,tension=0.25}{i3,i4}
 \fmf{plain,right=.33,tension=0.25}{i5,i6}
 \fmf{wiggly,right=.33,tension=0.25}{i2,i3}
  \fmf{wiggly,right=.33,tension=0.25}{i4,i5}
 \fmf{wiggly,right=.33,tension=0.25}{i6,i1}
 \end{fmfchar*}}
 \quad + \ldots
\end{equation}
or
\begin{eqnarray}
  \nonumber Z[1+u]/Z_0 &=& 1 +  \quad
    \parbox{10mm}{\begin{fmfchar*}(10,10)
 \fmfpen{thin}
 \fmfset{wiggly_len}{2mm}
 \fmftop{i1} \fmfbottom{o1}
 \fmfdot{i1}
 \fmfdot{o1}
 \fmf{plain,right,tension=0.5}{i1,o1}
 \fmf{wiggly,right,tension=0.5}{o1,i1}
 \end{fmfchar*}}
  \quad + \left\{\;
    \parbox{10mm}{\begin{fmfchar*}(10,10)
 \fmfpen{thin}
 \fmfset{wiggly_len}{2mm}
 \fmftop{i1} \fmfbottom{o1}
 \fmfdot{i1}
 \fmfdot{o1}
 \fmf{plain,right,tension=0.5}{i1,o1}
 \fmf{wiggly,right,tension=0.5}{o1,i1}
 \end{fmfchar*}}
 \; \;
    \parbox{10mm}{\begin{fmfchar*}(10,10)
 \fmfpen{thin}
 \fmfset{wiggly_len}{2mm}
 \fmftop{i1} \fmfbottom{o1}
 \fmfdot{i1}
 \fmfdot{o1}
 \fmf{plain,right,tension=0.5}{i1,o1}
 \fmf{wiggly,right,tension=0.5}{o1,i1}
 \end{fmfchar*}}
  \quad + \quad
    \parbox{10mm}{\begin{fmfchar*}(10,10)
 \fmfpen{thin}
 \fmfset{wiggly_len}{2mm}
 \fmfsurround{i1,i2,i3,i4}
 \fmfdot{i1,i2,i3,i4}
 \fmf{plain,right=.5,tension=0.35}{i1,i2}
 \fmf{plain,right=.5,tension=0.25}{i3,i4}
 \fmf{wiggly,right=.5,tension=0.25}{i2,i3}
 \fmf{wiggly,right=.5,tension=0.25}{i4,i1}
 \end{fmfchar*}}
   \; \right\} \\
 && + \left\{ \;
    \parbox{10mm}{\begin{fmfchar*}(10,10)
 \fmfpen{thin}
 \fmfset{wiggly_len}{2mm}
 \fmftop{i1} \fmfbottom{o1}
 \fmfdot{i1}
 \fmfdot{o1}
 \fmf{plain,right,tension=0.5}{i1,o1}
 \fmf{wiggly,right,tension=0.5}{o1,i1}
 \end{fmfchar*}}
 \; \;
    \parbox{10mm}{\begin{fmfchar*}(10,10)
 \fmfpen{thin}
 \fmfset{wiggly_len}{2mm}
 \fmftop{i1} \fmfbottom{o1}
 \fmfdot{i1}
 \fmfdot{o1}
 \fmf{plain,right,tension=0.5}{i1,o1}
 \fmf{wiggly,right,tension=0.5}{o1,i1}
 \end{fmfchar*}}
 \; \;
    \parbox{10mm}{\begin{fmfchar*}(10,10)
 \fmfpen{thin}
 \fmfset{wiggly_len}{2mm}
 \fmftop{i1} \fmfbottom{o1}
 \fmfdot{i1}
 \fmfdot{o1}
 \fmf{plain,right,tension=0.5}{i1,o1}
 \fmf{wiggly,right,tension=0.5}{o1,i1}
 \end{fmfchar*}}
 \quad + \quad  \ldots \right\} + \ldots
\end{eqnarray}
which includes for each internal point a summation, a minus sign for
each closed fermion loop and the factor 1/(order of invariance
group)  for each diagram. It is important for the symmetry, that in
the expansion of the Pfaffian the fermion lines carry no direction.
However, translating the diagrams one has to go through it in a
certain direction, which determines the sign of each element. If we
went through it in the opposite direction, all elements would obtain
the opposite sign, because they are skew-symmetric. The result is
the same. $Z_0$
is just given by $Z_0 = Z[1] = 1$. \\

\noindent Introducing the kernel ${\cal K}_N (z_1,z_2)$ from eq.
(\ref{54a}) and ${\cal F} (z_1,z_2)$ from eq. (\ref{39}) we
reinterpret the diagrams as
\begin{equation}\label{71}
{\cal K}_N (z_1,z_2) = {\cal K}_N (1,2) =\quad
\parbox{25mm}{\begin{fmfchar*}(15,10)
 \fmfpen{thin}
 \fmfset{wiggly_len}{2mm}
 \fmfleft{i1} \fmfright{o1}
 \fmflabel{$1$}{i1}
 \fmflabel{$2$}{o1}
 \fmf{plain}{i1,o1}
 \fmfdot{i1}
 \fmfdot{o1}
 \end{fmfchar*}}
\end{equation}
\begin{equation}\label{72}
- {\cal F} (z_1,z_2)  (u(z_1) + u(z_2) + u(z_1) u(z_2))= \quad
\parbox{25mm}{\begin{fmfchar*}(15,10)
 \fmfpen{thin}
 \fmfset{wiggly_len}{2mm}
 \fmfleft{i1} \fmfright{o1}
 \fmflabel{$1$}{i1}
 \fmflabel{$2$}{o1}
 \fmf{wiggly}{i1,o1}
 \fmfdot{i1}
 \fmfdot{o1}
 \end{fmfchar*}}
\end{equation}
and at each internal point $z$ we have now an integration over $d^2
z$. In an obvious notation we now produce all correlation functions
by functional derivatives  w.r.t. $u(z)$:
\begin{equation}
 \fl  \frac{\delta Z [1 + u]}{\delta u(z_1)} = \parbox{10mm}{\vspace*{2cm}}
 \parbox{8mm}{
 \begin{fmfchar*}(8,8)
 \fmfpen{thin}
 \fmfset{wiggly_len}{2mm}
 \fmfbottom{i1} \fmftop{o1}
 \fmflabel{$1$}{o1}
 \fmfdot{o1} \fmfdot{i1}
  \fmf{plain,left,tension=0.5}{i1,o1} \fmf{wiggly,left,tension=0.5}{o1,i1}
 \end{fmfchar*}}
  + \left\{ \;
 \parbox{10mm}{
 \begin{fmfchar*}(8,8)
 \fmfpen{thin}
 \fmfset{wiggly_len}{2mm}
 \fmfbottom{i1} \fmftop{o1}
 \fmflabel{$1$}{o1}
 \fmfdot{o1} \fmfdot{i1}
  \fmf{plain,left,tension=0.5}{i1,o1} \fmf{wiggly,left,tension=0.5}{o1,i1}
 \end{fmfchar*}}
 \;
    \parbox{8mm}{\begin{fmfchar*}(8,8)
 \fmfpen{thin}
 \fmfset{wiggly_len}{2mm}
 \fmftop{i1} \fmfbottom{o1}
 \fmfdot{i1}
 \fmfdot{o1}
 \fmf{plain,right,tension=0.5}{i1,o1}
 \fmf{wiggly,right,tension=0.5}{o1,i1}
 \end{fmfchar*}}
 \quad + \quad
    \parbox{8mm}{\begin{fmfchar*}(8,8)
 \fmfpen{thin}
 \fmfset{wiggly_len}{2mm}
 \fmfsurround{i1,i2,i3,i4}
 \fmfdot{i1,i2,i3,i4}
 \fmflabel{$1$}{i2}
 \fmf{plain,right=.5,tension=0.35}{i1,i2}
 \fmf{plain,right=.5,tension=0.25}{i3,i4}
 \fmf{wiggly,right=.5,tension=0.25}{i2,i3}
 \fmf{wiggly,right=.5,tension=0.25}{i4,i1}
 \end{fmfchar*}}
\; \right\}+\left\{
 \parbox{8mm}{
 \begin{fmfchar*}(8,8)
 \fmfpen{thin}
 \fmfset{wiggly_len}{2mm}
 \fmfbottom{i1} \fmftop{o1}
 \fmflabel{$1$}{o1}
 \fmfdot{o1} \fmfdot{i1}
  \fmf{plain,left,tension=0.5}{i1,o1} \fmf{wiggly,left,tension=0.5}{o1,i1}
 \end{fmfchar*}}
 \;
    \parbox{8mm}{\begin{fmfchar*}(8,8)
 \fmfpen{thin}
 \fmfset{wiggly_len}{2mm}
 \fmftop{i1} \fmfbottom{o1}
 \fmfdot{i1}
 \fmfdot{o1}
 \fmf{plain,right,tension=0.5}{i1,o1}
 \fmf{wiggly,right,tension=0.5}{o1,i1}
 \end{fmfchar*}}
 \;
    \parbox{8mm}{\begin{fmfchar*}(8,8)
 \fmfpen{thin}
 \fmfset{wiggly_len}{2mm}
 \fmftop{i1} \fmfbottom{o1}
 \fmfdot{i1}
 \fmfdot{o1}
 \fmf{plain,right,tension=0.5}{i1,o1}
 \fmf{wiggly,right,tension=0.5}{o1,i1}
 \end{fmfchar*}}
+\ldots \right\} + \ldots
\end{equation}
\vspace*{1cm}
\begin{eqnarray}
 \fl  \frac{\delta^2 Z [1 + u]}{\delta u(z_1) \delta u(z_2)} &=&
 \parbox{10mm}{
 \begin{fmfchar*}(10,10)
 \fmfpen{thin}
 \fmfset{wiggly_len}{2mm}
 \fmfbottom{i1} \fmftop{o1}
 \fmflabel{$1$}{o1}
 \fmflabel{$2$}{i1}
 \fmfdot{o1} \fmfdot{i1}
  \fmf{plain,left,tension=0.5}{i1,o1} \fmf{wiggly,left,tension=0.5}{o1,i1}
 \end{fmfchar*}}
 \; + \; \left\{ \;
 \parbox{10mm}{
 \begin{fmfchar*}(10,10)
 \fmfpen{thin}
 \fmfset{wiggly_len}{2mm}
 \fmfbottom{i1} \fmftop{o1}
 \fmflabel{$1$}{o1}
  \fmflabel{$2$}{i1}
 \fmfdot{o1} \fmfdot{i1}
  \fmf{plain,left,tension=0.5}{i1,o1} \fmf{wiggly,left,tension=0.5}{o1,i1}
 \end{fmfchar*}}
 \;
 \parbox{10mm}{
 \begin{fmfchar*}(10,10)
 \fmfpen{thin}
 \fmfset{wiggly_len}{2mm}
 \fmfbottom{i1} \fmftop{o1}
 \fmfdot{o1} \fmfdot{i1}
  \fmf{plain,left,tension=0.5}{i1,o1} \fmf{wiggly,left,tension=0.5}{o1,i1}
 \end{fmfchar*}}
 \; + \;
 \parbox{10mm}{
 \begin{fmfchar*}(10,10)
 \fmfpen{thin}
 \fmfset{wiggly_len}{2mm}
 \fmfbottom{i1} \fmftop{o1}
 \fmflabel{$1$}{o1}
 \fmfdot{o1} \fmfdot{i1}
  \fmf{plain,left,tension=0.5}{i1,o1} \fmf{wiggly,left,tension=0.5}{o1,i1}
 \end{fmfchar*}}
 \;
 \parbox{10mm}{
 \begin{fmfchar*}(10,10)
 \fmfpen{thin}
 \fmfset{wiggly_len}{2mm}
 \fmfbottom{i1} \fmftop{o1}
 \fmflabel{$2$}{o1}
 \fmfdot{o1} \fmfdot{i1}
  \fmf{plain,left,tension=0.5}{i1,o1} \fmf{wiggly,left,tension=0.5}{o1,i1}
 \end{fmfchar*}}
 \; + \right.\\
 &&  \parbox{10mm}{\vspace*{32mm}}
    \parbox{10mm}{\begin{fmfchar*}(10,10)
 \fmfpen{thin}
 \fmfset{wiggly_len}{2mm}
 \fmfsurround{i1,i2,i3,i4}
 \fmfdot{i1,i2,i3,i4}
 \fmflabel{$1$}{i1}
 \fmflabel{$2$}{i2}
 \fmf{plain,right=.5,tension=0.35}{i1,i2}
 \fmf{plain,right=.5,tension=0.25}{i3,i4}
 \fmf{wiggly,right=.5,tension=0.25}{i2,i3}
 \fmf{wiggly,right=.5,tension=0.25}{i4,i1}
 \end{fmfchar*}}
 \quad + \quad
    \parbox{10mm}{\begin{fmfchar*}(10,10)
 \fmfpen{thin}
 \fmfset{wiggly_len}{2mm}
 \fmfsurround{i1,i2,i3,i4}
 \fmfdot{i1,i2,i3,i4}
 \fmflabel{$1$}{i2}
 \fmflabel{$2$}{i4}
 \fmf{plain,right=.5,tension=0.35}{i1,i2}
 \fmf{plain,right=.5,tension=0.25}{i3,i4}
 \fmf{wiggly,right=.5,tension=0.25}{i2,i3}
 \fmf{wiggly,right=.5,tension=0.25}{i4,i1}
 \end{fmfchar*}}
\quad  + \quad
    \parbox{10mm}{\begin{fmfchar*}(10,10)
 \fmfpen{thin}
 \fmfset{wiggly_len}{2mm}
 \fmfsurround{i1,i2,i3,i4}
 \fmfdot{i1,i2,i3,i4}
 \fmflabel{$1$}{i2}
 \fmflabel{$2$}{i3}
 \fmf{plain,right=.5,tension=0.35}{i1,i2}
 \fmf{plain,right=.5,tension=0.25}{i3,i4}
 \fmf{wiggly,right=.5,tension=0.25}{i2,i3}
 \fmf{wiggly,right=.5,tension=0.25}{i4,i1}
 \end{fmfchar*}}
 \quad \Bigr\} + \ldots
\end{eqnarray}
And so on. A label $1$ means a functional derivative w.r.t. $u(z_1)$
and thus at this vertex appears no integration. At a wavy line one
can only differentiate twice and only at different vertices,
otherwise there is no contribution. To obtain all correlation
functions or $n$-point densities we have to put $u(z) \equiv 0$ at
the end. Then a lot of diagrams disappear. In the following diagrams
we reinterpret a wavy line as:
\begin{equation}\label{75}
\parbox{25mm}{\begin{fmfchar*}(15,10)
 \fmfpen{thin}
 \fmfset{wiggly_len}{2mm}
 \fmfleft{i1} \fmfright{o1}
 \fmflabel{$1$}{i1}
 \fmflabel{$2$}{o1}
 \fmf{wiggly}{i1,o1}
 \fmfdot{i1}
 \fmfdot{o1}
 \end{fmfchar*}}
   \Rightarrow - {\cal F} (z_1, z_2)
\end{equation}
and obtain
\begin{eqnarray}
\label{76}
  R_1 (z_1) &=& \quad \quad    \parbox{2mm}{\vspace{25mm}}
 \parbox{10mm}{
 \begin{fmfchar*}(10,10)
 \fmfpen{thin}
 \fmfset{wiggly_len}{2mm}
 \fmfbottom{i1} \fmftop{o1}
 \fmflabel{$1$}{o1}
 \fmfdot{o1} \fmfdot{i1}
  \fmf{plain,left,tension=0.5}{i1,o1} \fmf{wiggly,left,tension=0.5}{o1,i1}
 \end{fmfchar*}}
   \\
   \label{78a}
   \parbox{10mm}{\vspace*{2cm}}
  R_2(z_1,z_2) &=&
 \parbox{10mm}{
 \begin{fmfchar*}(10,10)
 \fmfpen{thin}
 \fmfset{wiggly_len}{2mm}
 \fmfbottom{i1} \fmftop{o1}
 \fmflabel{$1$}{o1}
 \fmflabel{$2$}{i1}
 \fmfdot{o1} \fmfdot{i1}
  \fmf{plain,left,tension=0.5}{i1,o1} \fmf{wiggly,left,tension=0.5}{o1,i1}
 \end{fmfchar*}}
 \; + \quad
 \parbox{10mm}{
 \begin{fmfchar*}(10,10)
 \fmfpen{thin}
 \fmfset{wiggly_len}{2mm}
 \fmfbottom{i1} \fmftop{o1}
 \fmflabel{$1$}{o1}
 \fmfdot{o1} \fmfdot{i1}
  \fmf{plain,left,tension=0.5}{i1,o1} \fmf{wiggly,left,tension=0.5}{o1,i1}
 \end{fmfchar*}}
 \;
 \parbox{10mm}{
 \begin{fmfchar*}(10,10)
 \fmfpen{thin}
 \fmfset{wiggly_len}{2mm}
 \fmfbottom{i1} \fmftop{o1}
 \fmflabel{$2$}{o1}
 \fmfdot{o1} \fmfdot{i1}
  \fmf{plain,left,tension=0.5}{i1,o1} \fmf{wiggly,left,tension=0.5}{o1,i1}
 \end{fmfchar*}}
 \quad + \quad
    \parbox{15mm}{\begin{fmfchar*}(10,10)
 \fmfpen{thin}
 \fmfset{wiggly_len}{2mm}
 \fmfsurround{i1,i2,i3,i4}
 \fmfdot{i1,i2,i3,i4}
 \fmflabel{$1$}{i2}
 \fmflabel{$2$}{i4}
 \fmf{plain,right=.5,tension=0.35}{i1,i2}
 \fmf{plain,right=.5,tension=0.25}{i3,i4}
 \fmf{wiggly,right=.5,tension=0.25}{i2,i3}
 \fmf{wiggly,right=.5,tension=0.25}{i4,i1}
 \end{fmfchar*}}
 + \quad
    \parbox{20mm}{\begin{fmfchar*}(10,10)
 \fmfpen{thin}
 \fmfset{wiggly_len}{2mm}
 \fmfsurround{i1,i2,i3,i4}
 \fmfdot{i1,i2,i3,i4}
 \fmflabel{$1$}{i2}
 \fmflabel{$2$}{i1}
 \fmf{plain,right=.5,tension=0.35}{i1,i2}
 \fmf{plain,right=.5,tension=0.25}{i3,i4}
 \fmf{wiggly,right=.5,tension=0.25}{i2,i3}
 \fmf{wiggly,right=.5,tension=0.25}{i4,i1}
 \end{fmfchar*}}
\end{eqnarray}
And so on. These diagrams have no longer an invariance group,
because the external vertices, which are not integrated over are
labeled and distinguishable. Here no wavy line is possible that has
only internal vertices. Differentiating a diagram, which contains (\ref{72}), once w.r.t.
$u(z_1)$ at $u \equiv 0$ yields a wavy line with one external vertex
 (1). Differentiating a diagram, which contains (\ref{72}),  w.r.t. $u(z_1)$ and $u(z_2)$ at $u \equiv 0$
 yields a wavy line with two external vertices (1,2). \\
We could also calculate the cluster functions which are in each
order only the connected diagrams, i.e. here the one-loop diagrams.\\

\noindent We will see that in the case of odd $N$ the diagrams are
slightly modified. There is an additional graphical element for the
correlation functions (independent of the direction going through)
\begin{equation}\label{78}
\parbox{25mm}{\begin{fmfchar*}(15,10)
 \fmfpen{thin}
 \fmfset{wiggly_len}{2mm}
 \fmfleft{i1} \fmfright{o1}
 \fmflabel{$1$}{i1}
 \fmflabel{$2$}{o1}
 \fmf{plain}{i1,v1}
 \fmfv{decor.shape=cross,decoration.size=2.5mm}{v1}
 \fmf{wiggly}{v1,o1}
 \fmfdot{i1}
 \fmfdot{o1}
 \end{fmfchar*}}
   = (-) \frac{z_1^{N-1}}{2^{N/2} \Gamma(N/2)} \cdot
     {\rm e}^{-x_2^2/2} \delta(y_2) \Theta(N \mbox{ odd})
\end{equation}
The cross corresponds to the additional artificial Grassmannian and
therefore can only appear once in each diagram. The result for the
correlation functions is then
\begin{eqnarray}
\label{84a}
  R_1(z_1) &=& \parbox{10mm}{\vspace*{2.5cm}}
 \parbox{10mm}{
 \begin{fmfchar*}(10,10)
 \fmfpen{thin}
 \fmfset{wiggly_len}{2mm}
 \fmfbottom{i1} \fmftop{o1}
 \fmflabel{$1$}{o1}
 \fmfdot{o1} \fmfdot{i1}
  \fmf{plain,left,tension=0.5}{i1,o1} \fmf{wiggly,left,tension=0.5}{o1,i1}
 \end{fmfchar*}}
 \; + \;
 \parbox{10mm}{
 \begin{fmfchar*}(10,10)
 \fmfpen{thin}
 \fmfset{wiggly_len}{2mm}
 \fmfbottom{i1} \fmftop{o1}
 \fmflabel{$1$}{o1}
 \fmfdot{o1}
 \fmfv{decor.shape=cross,decoration.size=2.5mm}{i1}
  \fmf{plain,left,tension=0.5}{i1,o1} \fmf{wiggly,left,tension=0.5}{o1,i1}
 \end{fmfchar*}}
  \\
 \nonumber R_2 (z_1,z_2) &=& \parbox{10mm}{\vspace*{2.5cm}}
 \parbox{10mm}{
 \begin{fmfchar*}(10,10)
 \fmfpen{thin}
 \fmfset{wiggly_len}{2mm}
 \fmfbottom{i1} \fmftop{o1}
 \fmflabel{$1$}{o1}
  \fmflabel{$2$}{i1}
 \fmfdot{o1} \fmfdot{i1}
  \fmf{plain,left,tension=0.5}{i1,o1} \fmf{wiggly,left,tension=0.5}{o1,i1}
 \end{fmfchar*}}
 \; + \;
 \parbox{10mm}{
 \begin{fmfchar*}(10,10)
 \fmfpen{thin}
 \fmfset{wiggly_len}{2mm}
 \fmfbottom{i1} \fmftop{o1}
 \fmflabel{$1$}{o1}
 \fmfdot{o1} \fmfdot{i1}
  \fmf{plain,left,tension=0.5}{i1,o1} \fmf{wiggly,left,tension=0.5}{o1,i1}
 \end{fmfchar*}}
 \;
 \parbox{10mm}{
 \begin{fmfchar*}(10,10)
 \fmfpen{thin}
 \fmfset{wiggly_len}{2mm}
 \fmfbottom{i1} \fmftop{o1}
 \fmflabel{$2$}{o1}
 \fmfdot{o1} \fmfdot{i1}
  \fmf{plain,left,tension=0.5}{i1,o1} \fmf{wiggly,left,tension=0.5}{o1,i1}
 \end{fmfchar*}}
 \; + \;
 \parbox{10mm}{
 \begin{fmfchar*}(10,10)
 \fmfpen{thin}
 \fmfset{wiggly_len}{2mm}
 \fmfbottom{i1} \fmftop{o1}
 \fmflabel{$1$}{o1}
 \fmfdot{o1}
 \fmfv{decor.shape=cross,decoration.size=2.5mm}{i1}
  \fmf{plain,left,tension=0.5}{i1,o1} \fmf{wiggly,left,tension=0.5}{o1,i1}
 \end{fmfchar*}}
 \;
 \parbox{10mm}{
 \begin{fmfchar*}(10,10)
 \fmfpen{thin}
 \fmfset{wiggly_len}{2mm}
 \fmfbottom{i1} \fmftop{o1}
 \fmflabel{$2$}{o1}
 \fmfdot{o1} \fmfdot{i1}
  \fmf{plain,left,tension=0.5}{i1,o1} \fmf{wiggly,left,tension=0.5}{o1,i1}
 \end{fmfchar*}}
 \; + \;
 \parbox{10mm}{
 \begin{fmfchar*}(10,10)
 \fmfpen{thin}
 \fmfset{wiggly_len}{2mm}
 \fmfbottom{i1} \fmftop{o1}
 \fmflabel{$1$}{o1}
 \fmfdot{o1} \fmfdot{i1}
  \fmf{plain,left,tension=0.5}{i1,o1} \fmf{wiggly,left,tension=0.5}{o1,i1}
 \end{fmfchar*}}
 \;
 \parbox{10mm}{
 \begin{fmfchar*}(10,10)
 \fmfpen{thin}
 \fmfset{wiggly_len}{2mm}
 \fmfbottom{i1} \fmftop{o1}
 \fmflabel{$2$}{o1}
 \fmfv{decor.shape=cross,decoration.size=2.5mm}{i1}
 \fmfdot{o1}
  \fmf{plain,left,tension=0.5}{i1,o1} \fmf{wiggly,left,tension=0.5}{o1,i1}
 \end{fmfchar*}}
 \\
 \nonumber
 \label{80}  &&   + \quad  \parbox{10mm}{\vspace*{2.5cm}}
    \parbox{10mm}{\begin{fmfchar*}(10,10)
 \fmfpen{thin}
 \fmfset{wiggly_len}{2mm}
 \fmfsurround{i1,i2,i3,i4}
 \fmfdot{i1,i2,i3,i4}
 \fmflabel{$1$}{i2}
 \fmflabel{$2$}{i4}
 \fmf{plain,right=.5,tension=0.35}{i1,i2}
 \fmf{plain,right=.5,tension=0.25}{i3,i4}
 \fmf{wiggly,right=.5,tension=0.25}{i2,i3}
 \fmf{wiggly,right=.5,tension=0.25}{i4,i1}
 \end{fmfchar*}}
\quad  + \quad
    \parbox{10mm}{\begin{fmfchar*}(10,10)
 \fmfpen{thin}
 \fmfset{wiggly_len}{2mm}
 \fmfsurround{i1,i2,i3,i4}
 \fmfdot{i2,i4,i3}
 \fmflabel{$1$}{i2}
 \fmflabel{$2$}{i4}
 \fmfv{decor.shape=cross,decoration.size=2.5mm}{i1}
 \fmf{plain,right=.5,tension=0.35}{i1,i2}
 \fmf{plain,right=.5,tension=0.25}{i3,i4}
 \fmf{wiggly,right=.5,tension=0.25}{i2,i3}
 \fmf{wiggly,right=.5,tension=0.25}{i4,i1}
 \end{fmfchar*}}
\quad + \quad
    \parbox{10mm}{\begin{fmfchar*}(10,10)
 \fmfpen{thin}
 \fmfset{wiggly_len}{2mm}
 \fmfsurround{i1,i2,i3,i4}
 \fmfdot{i1,i2,i4}
 \fmflabel{$1$}{i2}
 \fmflabel{$2$}{i4}
 \fmfv{decor.shape=cross,decoration.size=2.5mm}{i3}
 \fmf{plain,right=.5,tension=0.35}{i1,i2}
 \fmf{plain,right=.5,tension=0.25}{i3,i4}
 \fmf{wiggly,right=.5,tension=0.25}{i2,i3}
 \fmf{wiggly,right=.5,tension=0.25}{i4,i1}
 \end{fmfchar*}}
\\
\label{85a} && \parbox{10mm}{\vspace*{2.5cm}} + \quad
    \parbox{10mm}{\begin{fmfchar*}(10,10)
 \fmfpen{thin}
 \fmfset{wiggly_len}{2mm}
 \fmfsurround{i1,i2,i3,i4}
 \fmfdot{i1,i2,i3,i4}
 \fmflabel{$1$}{i3}
 \fmflabel{$2$}{i4}
 \fmf{plain,right=.5,tension=0.35}{i1,i2}
 \fmf{plain,right=.5,tension=0.25}{i3,i4}
 \fmf{wiggly,right=.5,tension=0.25}{i2,i3}
 \fmf{wiggly,right=.5,tension=0.25}{i4,i1}
 \end{fmfchar*}}
\quad  + \quad
    \parbox{10mm}{\begin{fmfchar*}(10,10)
 \fmfpen{thin}
 \fmfset{wiggly_len}{2mm}
 \fmfsurround{i1,i2,i3,i4}
 \fmfdot{i1,i3,i4}
 \fmflabel{$1$}{i3}
 \fmflabel{$2$}{i4}
 \fmfv{decor.shape=cross,decoration.size=2.5mm}{i2}
 \fmf{plain,right=.5,tension=0.35}{i1,i2}
 \fmf{plain,right=.5,tension=0.25}{i3,i4}
 \fmf{wiggly,right=.5,tension=0.25}{i2,i3}
 \fmf{wiggly,right=.5,tension=0.25}{i4,i1}
 \end{fmfchar*}}
\quad  + \quad
    \parbox{10mm}{\begin{fmfchar*}(10,10)
 \fmfpen{thin}
 \fmfset{wiggly_len}{2mm}
 \fmfsurround{i1,i2,i3,i4}
 \fmfdot{i1,i2,i4}
 \fmflabel{$1$}{i1}
 \fmflabel{$2$}{i2}
 \fmfv{decor.shape=cross,decoration.size=2.5mm}{i3}
 \fmf{plain,right=.5,tension=0.35}{i1,i2}
 \fmf{plain,right=.5,tension=0.25}{i3,i4}
 \fmf{wiggly,right=.5,tension=0.25}{i2,i3}
 \fmf{wiggly,right=.5,tension=0.25}{i4,i1}
 \end{fmfchar*}}
\end{eqnarray}
And so on. The terms with the crosses, which are only present if $N$
is odd, are obtained by labeling in all possible ways one internal
vertex in the original diagrams by a cross. One has still  to keep in
mind that closing a loop yields an additional minus sign.
\subsection{Correlation functions, $N$ even}
Let us translate the diagrams again (for even $N$): eq. (\ref{76})
leads to eq. (\ref{48}) and eq. (\ref{78a}) to
\begin{eqnarray}
 \nonumber R_2(z_1,z_2) &=& {\cal F} (z_1,z_2)  {\cal K}_N(z_2,z_1) + \int d^2 z_3  d^2 z_4  \{ \\
\nonumber &&+ {\cal F} (z_1,z_3) \, {\cal K}_N (z_3, z_1) \,{\cal F} (z_2,z_4) \,{\cal K}_N(z_4, z_2) \\
\nonumber &&- {\cal F} (z_1,z_3)\, {\cal K}_N(z_3, z_2) \,{\cal F} (z_2,z_4) \,{\cal K}_N(z_4, z_1) \\
&&- {\cal F} (z_1,z_3)\, {\cal K}_N(z_3, z_4)\, {\cal F} (z_4,z_2)\,
{\cal K}_N(z_2, z_1) \}
\end{eqnarray}
And so on. Now we know how to generate general $n$-point densities.
We observe that at the external vertices there are no integrations.
Thus the diagrams are cut into factors which are special
diagrammatic elements. These are used to build for an $n$-point
density closed loops with $n$ external vertices, using the rule that
a wavy line can only be linked  with a straight line, as is done in
eq. (\ref{85a}). In the following we will show that the result is
just what is called a quaternion determinant of a self-dual $2n
\times 2n$ matrix ($k,l = 1,2, \ldots, n$):
\begin{equation}\label{82}
\fl     R_n(z_1,z_2,\ldots, z_n) = (-1)^n \mbox{ qdet } \left(
\begin{array}{cc}
\parbox{15mm}{\begin{fmfchar*}(15,10)
 \fmfpen{thin}
 \fmfset{wiggly_len}{2mm}
 \fmfleft{i1} \fmfright{o1}
 \fmflabel{$k$}{i1}
 \fmflabel{$l$}{o1}
 \fmf{plain}{i1,v1}
 \fmf{wiggly}{v1,o1}
 \fmfdot{i1}
  \fmfdot{v1}
 \fmfdot{o1}
 \end{fmfchar*}}
 &
\parbox{10mm}{\begin{fmfchar*}(10,10)
 \fmfpen{thin}
 \fmfset{wiggly_len}{2mm}
 \fmfleft{i1} \fmfright{o1}
 \fmflabel{$k$}{i1}
 \fmflabel{$l$}{o1}
 \fmf{plain}{i1,o1}
 \fmfdot{i1}
 \fmfdot{o1}
 \end{fmfchar*}}
 \\
(\quad \parbox{10mm}{\begin{fmfchar*}(10,10)
 \fmfpen{thin}
 \fmfset{wiggly_len}{2mm}
 \fmfleft{i1} \fmfright{o1}
 \fmflabel{$k$}{i1}
 \fmflabel{$l$}{o1}
 \fmf{wiggly}{i1,o1}
 \fmfdot{i1}
 \fmfdot{o1}
 \end{fmfchar*}}
  \quad + \quad
\parbox{15mm}{\begin{fmfchar*}(15,10)
 \fmfpen{thin}
 \fmfset{wiggly_len}{2mm}
 \fmfleft{i1} \fmfright{o1}
 \fmflabel{$k$}{i1}
 \fmflabel{$l$}{o1}
 \fmf{wiggly}{i1,v1}
 \fmf{plain}{v1,v2}
 \fmf{wiggly}{v2,o1}
 \fmfdot{i1}
 \fmfdot{v1}
 \fmfdot{v2}
 \fmfdot{o1}
 \end{fmfchar*}} \quad )
 & \quad
\parbox{10mm}{\begin{fmfchar*}(10,10)
 \fmfpen{thin}
 \fmfset{wiggly_len}{2mm}
 \fmfleft{i1} \fmfright{o1}
 \fmflabel{$k$}{i1}
 \fmflabel{$l$}{o1}
\fmf{wiggly}{i1,v1}
 \fmf{plain}{v1,o1}
 \fmfdot{i1}
 \fmfdot{v1}
 \fmfdot{o1}
 \end{fmfchar*}}
 \quad
\end{array} \right)
\end{equation}
The entries are the appearing graphical elements. The $(-1)^n$
reminds us that we have to take into account a minus sign if we
close a loop. The sign of the quaternion determinant is defined in
such a way that in the expansion of $R_n(z_1,z_2,\ldots,z_n)$ there
appears the positive term $R_1(z_1) \cdot R_1(z_2) \cdot \ldots
\cdot R_1(z_n)$ coming from the diagonal elements, which gives the
behaviour for large separation. We will see that the quaternion
determinant is related to a Pfaffian. Since by construction
integration of $R_n$ over $d^2 z_n$ leads to $(N - n +1 )R_{n-1}$,
eq. (\ref{82}) implies an integration theorem \cite{Akemann07} for
this type of Pfaffians. This however includes here all $\delta$-type
contributions to $R_n$ ($\delta$-functions for real eigenvalues and
complex conjugate pairs). To separate all these  terms may still
require some combinatorial analysis.
\subsection{Expansion of a Pfaffian}
Let us consider
\begin{equation}\label{83}
    Z(\varepsilon) = \int d\eta_1^{*} \ldots d\eta_{2M}^{*} \; \exp \left(
    - \frac{1}{2} \sum_{k,l}^{1\ldots 2M} \eta_k^{*} (A_{kl} + C_{kl}) \eta_{l}^{*} \right)
\end{equation}
now with $A_{kl} = \varepsilon J_{kl}$ and $J = {\;0\; \; 1 \choose
-1 \; \,0} $ with $M \times M$ entries. Obviously $ \mbox{Pfaff }
(C_{kl}) = \lim_{ \varepsilon \rightarrow 0} Z( \varepsilon)$. Now
we expand $Z(\varepsilon)$ in powers of $C$ using the fermionic Wick
theorem with $A^{-1}_{kl} = - \frac{1}{\varepsilon} J_{kl}$. Since
$Z_0 = (-1)^{M(M-1)/2} \varepsilon^M$ in the limit $\varepsilon
\rightarrow 0$ only terms with power $M$ of $A^{-1}$ survive, higher
powers do not occur. On the other hand $\langle \eta_k^{*}
\eta_l^{*} \rangle_{0} = - \frac{1}{\varepsilon} J_{kl}$. Thus we
have
\begin{eqnarray}
 \fl  \nonumber (-1)^{M(M-1)/2} \mbox{ Pfaff }(C) = \mbox{ sum of all diagrams of order }M \\
\nonumber   = \Bigl\{ \underbrace{  \parbox{2mm}{\vspace{15mm}}
 \parbox{10mm}{
 \begin{fmfchar*}(10,10)
 \fmfpen{thin}
 \fmfset{wiggly_len}{2mm}
 \fmfbottom{i1} \fmftop{o1}
 \fmfdot{o1} \fmfdot{i1}
  \fmf{plain,left,tension=0.5}{i1,o1} \fmf{wiggly,left,tension=0.5}{o1,i1}
 \end{fmfchar*}}
 \;
 \parbox{10mm}{
 \begin{fmfchar*}(10,10)
 \fmfpen{thin}
 \fmfset{wiggly_len}{2mm}
 \fmfbottom{i1} \fmftop{o1}
 \fmfdot{o1} \fmfdot{i1}
  \fmf{plain,left,tension=0.5}{i1,o1} \fmf{wiggly,left,tension=0.5}{o1,i1}
 \end{fmfchar*}}
 \ldots
 \parbox{10mm}{
 \begin{fmfchar*}(10,10)
 \fmfpen{thin}
 \fmfset{wiggly_len}{2mm}
 \fmfbottom{i1} \fmftop{o1}
 \fmfdot{o1} \fmfdot{i1}
  \fmf{plain,left,tension=0.5}{i1,o1} \fmf{wiggly,left,tension=0.5}{o1,i1}
 \end{fmfchar*}}
 }_{\mbox{$M$  times }} \Bigr\} \quad + \quad \Bigl\{
    \parbox{10mm}{\begin{fmfchar*}(10,10)
 \fmfpen{thin}
 \fmfset{wiggly_len}{2mm}
 \fmfsurround{i1,i2,i3,i4}
 \fmfdot{i1,i2,i3,i4}
 \fmf{plain,right=.5,tension=0.35}{i1,i2}
 \fmf{plain,right=.5,tension=0.25}{i3,i4}
 \fmf{wiggly,right=.5,tension=0.25}{i2,i3}
 \fmf{wiggly,right=.5,tension=0.25}{i4,i1}
 \end{fmfchar*}}
\quad \underbrace{ \parbox{2mm}{\vspace{15mm}}
 \parbox{10mm}{
 \begin{fmfchar*}(10,10)
 \fmfpen{thin}
 \fmfset{wiggly_len}{2mm}
 \fmfbottom{i1} \fmftop{o1}
 \fmfdot{o1} \fmfdot{i1}
  \fmf{plain,left,tension=0.5}{i1,o1} \fmf{wiggly,left,tension=0.5}{o1,i1}
 \end{fmfchar*}}
 \;
 \parbox{10mm}{
 \begin{fmfchar*}(10,10)
 \fmfpen{thin}
 \fmfset{wiggly_len}{2mm}
 \fmfbottom{i1} \fmftop{o1}
 \fmfdot{o1} \fmfdot{i1}
  \fmf{plain,left,tension=0.5}{i1,o1} \fmf{wiggly,left,tension=0.5}{o1,i1}
 \end{fmfchar*}}
 \ldots
 \parbox{10mm}{
 \begin{fmfchar*}(10,10)
 \fmfpen{thin}
 \fmfset{wiggly_len}{2mm}
 \fmfbottom{i1} \fmftop{o1}
 \fmfdot{o1} \fmfdot{i1}
  \fmf{plain,left,tension=0.5}{i1,o1} \fmf{wiggly,left,tension=0.5}{o1,i1}
 \end{fmfchar*}}
 }_{\mbox{ $(M-2)$ times}} \Bigr\} + \ldots
 \\
 \label{84}  = (-1)^M \mbox{ qdet } (J\cdot C) = (-1)^M \mbox{ qdet }
 ( \quad
\parbox{15mm}{\begin{fmfchar*}(15,10)
 \fmfpen{thin}
 \fmfset{wiggly_len}{2mm}
 \fmfleft{i1} \fmfright{o1}
 \fmf{plain}{i1,v1}
 \fmf{wiggly}{v1,o1}
 \fmfdot{i1}
 \fmfdot{v1}
 \fmfdot{o1}
 \end{fmfchar*}}
\quad  )
\end{eqnarray}
with $ 
\quad \parbox{15mm}{\begin{fmfchar*}(15,10)
 \fmfpen{thin}
 \fmfset{wiggly_len}{2mm}
 \fmfleft{i1} \fmfright{o1}
 \fmf{plain}{i1,o1}
 \fmfdot{i1}
 \fmfdot{o1}
 \end{fmfchar*}}
   \; = - J$, $ 
\parbox{15mm}{\begin{fmfchar*}(15,10)
 \fmfpen{thin}
 \fmfset{wiggly_len}{2mm}
 \fmfleft{i1} \fmfright{o1}
 \fmf{wiggly}{i1,o1}
 \fmfdot{i1}
 \fmfdot{o1}
 \end{fmfchar*}}
 \;  = -C$.\\
We see that only the matrix $JC$ appears which is selfdual (i.e.
$J(JC)^T J^T = JC$ if $C = - C^T$). We also may replace the
expansion of the Pfaffian by an indexed diagram expansion in which
all external vertices are different. To see this formally let
$C_{kl} \rightarrow \lambda_k C_{kl} \lambda_{l}$ with  $\lambda_k =
1$ for $k = M+1, M+2, \ldots , 2M$ and differentiate $\mbox{Pfaff
}(\lambda C \lambda)$ w.r.t. $\lambda_1,\lambda_2,\ldots,\lambda_M$.
This produces  the indexed diagram expansion and since $\mbox{Pfaff
} (\lambda C \lambda) = \mbox{ Pfaff } (C) \cdot \lambda_1 \lambda_2
\cdots \lambda_M$ the result is the same. This proves our claim
(\ref{82}). Note that if one uses a different definition of the
Pfaffian with a different order of Grassmannians the factor
$(-1)^{M(M-1)/2}$ in (\ref{84}) may be canceled.
\subsection{Correlation functions, $N$ odd}
Let us recall for odd $N$ the diagrammatic expansion on the level
where the vertices carry the number $k$ of Grassmannian
$\eta_k^{*}$. Then again
\begin{eqnarray}
 \nonumber   \langle \eta_k^{*}
\eta_l^{*} \rangle_{0} &=& B_{kl}^{-1} = \quad
\parbox{25mm}{\begin{fmfchar*}(15,10)
 \fmfpen{thin}
 \fmfset{wiggly_len}{2mm}
 \fmfleft{i1} \fmfright{o1}
 \fmflabel{$k$}{i1}
 \fmflabel{$l$}{o1}
 \fmf{plain}{i1,o1}
 \fmfdot{i1}
 \fmfdot{o1}
 \end{fmfchar*}}
 \\
 && -C_{kl} = \quad
\parbox{25mm}{\begin{fmfchar*}(15,10)
 \fmfpen{thin}
 \fmfset{wiggly_len}{2mm}
 \fmfleft{i1} \fmfright{o1}
 \fmflabel{$k$}{i1}
 \fmflabel{$l$}{o1}
 \fmf{wiggly}{i1,o1}
 \fmfdot{i1}
 \fmfdot{o1}
 \end{fmfchar*}}
\end{eqnarray}
with $k,l = 1,2, \ldots , N$ and  there are additional graphical
elements
\begin{eqnarray}
 \nonumber   \langle \eta_{N+1}^{*}
\eta_k^{*} \rangle_{0} = B_{N+1,k}^{-1} &=& \quad
\parbox{25mm}{\begin{fmfchar*}(15,10)
 \fmfpen{thin}
 \fmfset{wiggly_len}{2mm}
 \fmfleft{i1} \fmfright{o1}
 \fmfv{decor.shape=cross,decoration.size=2.5mm}{i1}
 \fmflabel{$k$}{o1}
 \fmf{plain}{i1,o1}
 \fmfdot{o1}
 \end{fmfchar*}}
 \\
- \int dx \, {\rm e}^{-x^2/2} x^{l} u(x)   &=& \quad
\parbox{25mm}{\begin{fmfchar*}(15,10)
 \fmfpen{thin}
 \fmfset{wiggly_len}{2mm}
 \fmfleft{i1} \fmfright{o1}
 \fmflabel{$l$}{i1}
 \fmf{wiggly}{i1,o1}
   \fmfv{decor.shape=cross,decoration.size=2.5mm}{o1}
 \fmfdot{i1}
 \end{fmfchar*}}
\end{eqnarray}
We have from eq. (\ref{63}) $B_{N+1,k}^{-1} = \delta_{kN}/2^{N/2}
\Gamma( N/2) = - B^{-1}_{k,N+1}$ and find that in the diagrammatic
expansion only the combination
\begin{equation}\label{87}
\parbox{25mm}{\begin{fmfchar*}(15,10)
 \fmfpen{thin}
 \fmfset{wiggly_len}{2mm}
 \fmfleft{i1} \fmfright{o1}
 \fmflabel{$k$}{i1}
 \fmflabel{$l$}{o1}
 \fmf{plain}{i1,v1}
 \fmf{wiggly}{v1,o1}
   \fmfv{decor.shape=cross,decoration.size=2.5mm}{v1}
 \fmfdot{i1}
 \fmfdot{o1}
 \end{fmfchar*}}
  = - \frac{\delta_{k,N}}{2^{N/2} \Gamma(N/2)} \cdot \int dx \, {\rm e}^{-x^2/2} x^{l-1} u(x)
\end{equation}
occurs. In this diagram it is not important in which direction one
goes through it: the result is the same. Thus one obtains the
additional diagrams:
\begin{equation}\label{88}
\fl  \parbox{10mm}{
 \begin{fmfchar*}(10,10)
 \fmfpen{thin}
 \fmfset{wiggly_len}{2mm}
 \fmfbottom{i1} \fmftop{o1}
 \fmfv{decor.shape=cross,decoration.size=2.5mm}{i1}
 \fmfdot{o1}
  \fmf{plain,left,tension=0.5}{i1,o1} \fmf{wiggly,left,tension=0.5}{o1,i1}
 \end{fmfchar*}}
 \quad + \quad \Bigl\{\;
    \parbox{10mm}{\begin{fmfchar*}(10,10)
 \fmfpen{thin}
 \fmfset{wiggly_len}{2mm}
 \fmfsurround{i1,i2,i3,i4}
 \fmfdot{i1,i2,i4}
 \fmfv{decor.shape=cross,decoration.size=2.5mm}{i3}
 \fmf{plain,right=.5,tension=0.35}{i1,i2}
 \fmf{plain,right=.5,tension=0.25}{i3,i4}
 \fmf{wiggly,right=.5,tension=0.25}{i2,i3}
 \fmf{wiggly,right=.5,tension=0.25}{i4,i1}
 \end{fmfchar*}}
\quad + \quad
 \parbox{10mm}{
 \begin{fmfchar*}(10,10)
 \fmfpen{thin}
 \fmfset{wiggly_len}{2mm}
 \fmfbottom{i1} \fmftop{o1}
 \fmfv{decor.shape=cross,decoration.size=2.5mm}{i1}
 \fmfdot{o1}
  \fmf{plain,left,tension=0.5}{i1,o1} \fmf{wiggly,left,tension=0.5}{o1,i1}
 \end{fmfchar*}}
 \;
 \parbox{10mm}{
 \begin{fmfchar*}(10,10)
 \fmfpen{thin}
 \fmfset{wiggly_len}{2mm}
 \fmfbottom{i1} \fmftop{o1}
 \fmfdot{o1}
 \fmfdot{i1}
  \fmf{plain,left,tension=0.5}{i1,o1} \fmf{wiggly,left,tension=0.5}{o1,i1}
 \end{fmfchar*}}
 \; \Bigr\}  \quad  + \quad \Bigl\{ \;
 \parbox{10mm}{
 \begin{fmfchar*}(10,10)
 \fmfpen{thin}
 \fmfset{wiggly_len}{2mm}
 \fmfbottom{i1} \fmftop{o1}
 \fmfv{decor.shape=cross,decoration.size=2.5mm}{i1}
 \fmfdot{o1}
  \fmf{plain,left,tension=0.5}{i1,o1} \fmf{wiggly,left,tension=0.5}{o1,i1}
 \end{fmfchar*}}
 \;
 \parbox{10mm}{
 \begin{fmfchar*}(10,10)
 \fmfpen{thin}
 \fmfset{wiggly_len}{2mm}
 \fmfbottom{i1} \fmftop{o1}
 \fmfdot{o1}
 \fmfdot{i1}
  \fmf{plain,left,tension=0.5}{i1,o1} \fmf{wiggly,left,tension=0.5}{o1,i1}
 \end{fmfchar*}}
 \;
 \parbox{10mm}{
 \begin{fmfchar*}(10,10)
 \fmfpen{thin}
 \fmfset{wiggly_len}{2mm}
 \fmfbottom{i1} \fmftop{o1}
 \fmfdot{o1}
 \fmfdot{i1}
  \fmf{plain,left,tension=0.5}{i1,o1} \fmf{wiggly,left,tension=0.5}{o1,i1}
 \end{fmfchar*}}
 \quad + \ldots \Bigr\} + \ldots
\end{equation}
Only diagrams with one cross are needed. At each vertex without
cross there appears a $k$-summation  over $z^{k-1}$. \\

\noindent Now we again may reinterpret the diagrams as:
\begin{eqnarray}
\nonumber 
\parbox{25mm}{\begin{fmfchar*}(15,10)
 \fmfpen{thin}
 \fmfset{wiggly_len}{2mm}
 \fmfleft{i1} \fmfright{o1}
 \fmflabel{$1$}{i1}
 \fmflabel{$2$}{o1}
 \fmf{plain}{i1,o1}
 \fmfdot{i1}
 \fmfdot{o1}
 \end{fmfchar*}}
  &=& {\cal K}_N (1,2) \\
\nonumber 
\parbox{25mm}{\begin{fmfchar*}(15,10)
 \fmfpen{thin}
 \fmfset{wiggly_len}{2mm}
 \fmfleft{i1} \fmfright{o1}
 \fmflabel{$1$}{i1}
 \fmflabel{$2$}{o1}
 \fmf{wiggly}{i1,o1}
 \fmfdot{i1}
 \fmfdot{o1}
 \end{fmfchar*}}
  &=& - {\cal F} (1,2) (u(1) + u(2) + u(1)u(2)) \\
\parbox{25mm}{\begin{fmfchar*}(15,10)
 \fmfpen{thin}
 \fmfset{wiggly_len}{2mm}
 \fmfleft{i1} \fmfright{o1}
 \fmflabel{$1$}{i1}
 \fmflabel{$2$}{o1}
 \fmf{wiggly}{i1,v1}
 \fmf{plain}{v1,o1}
 \fmfv{decor.shape=cross,decoration.size=2.5mm}{v1}
 \fmfdot{i1}
 \fmfdot{o1}
 \end{fmfchar*}}
 &=& -u(1) {\rm e}^{- x_1^2/2} \delta (y_1) z_2^{N-1}/2^{N/2} \Gamma(N/2)
\end{eqnarray}
with integration at internal vertices. Remember that a closed loop
yields a factor $(-1)$. Finally we differentiate the diagrams w.r.t.
$u(z)$ several times and put then  $u(z) \equiv 0$ to obtain all
correlation functions. Note that there  is no differentiation
possible at a cross. The result is the correlations (\ref{84a},
\ref{85a}) and
corresponding higher orders.\\

\noindent The general formula for the $n$-point densities may again
be written as a quaternion determinant with $k,l = 1, 2, \ldots, n$:
\begin{equation}\label{90}
 \fl    R_n (z_1,\ldots, z_n) = (-1)^n \mbox{ qdet} \left( \begin{array}{cc}
 \quad \parbox{10mm}{\begin{fmfchar*}(10,10)
 \fmfpen{thin}
 \fmfset{wiggly_len}{2mm}
 \fmfleft{i1} \fmfright{o1}
 \fmflabel{$(k$}{i1}
 \fmflabel{$l$}{o1}
 \fmf{plain}{i1,v1}
 \fmf{wiggly}{v1,o1}
 \fmfdot{i1}
 \fmfdot{o1}
 \fmfdot{v1}
 \end{fmfchar*}}
 \quad + \quad
  \parbox{10mm}{\begin{fmfchar*}(10,10)
 \fmfpen{thin}
 \fmfset{wiggly_len}{2mm}
 \fmfleft{i1} \fmfright{o1}
 \fmflabel{$k$}{i1}
 \fmflabel{$l)$}{o1}
 \fmf{plain}{i1,v1}
 \fmf{wiggly}{v1,o1}
 \fmfdot{i1}
 \fmfv{decor.shape=cross,decoration.size=2.5mm}{v1}
 \fmfdot{o1}
 \end{fmfchar*}}
 \quad &
\parbox{10mm}{\begin{fmfchar*}(10,10)
 \fmfpen{thin}
 \fmfset{wiggly_len}{2mm}
 \fmfleft{i1} \fmfright{o1}
 \fmflabel{$k$}{i1}
 \fmflabel{$l$}{o1}
 \fmf{plain}{i1,o1}
 \fmfdot{i1}
 \fmfdot{o1}
 \end{fmfchar*}}
 \\
\parbox{10mm}{\begin{fmfchar*}(10,10)
 \fmfpen{thin}
 \fmfset{wiggly_len}{2mm}
 \fmfleft{i1} \fmfright{o1}
 \fmflabel{$k$}{i1}
 \fmflabel{$l$}{o1}
 \fmf{dbl_wiggly}{i1,o1}
 \fmfdot{i1}
 \fmfdot{o1}
 \end{fmfchar*}}
 & \quad
\parbox{10mm}{\begin{fmfchar*}(10,10)
 \fmfpen{thin}
 \fmfset{wiggly_len}{2mm}
 \fmfleft{i1} \fmfright{o1}
 \fmflabel{$(k$}{i1}
 \fmflabel{$l$}{o1}
 \fmf{wiggly}{i1,v1}
 \fmf{plain}{v1,o1}
 \fmfdot{i1}
 \fmfdot{o1}
 \fmfdot{v1}
 \end{fmfchar*}}
 \quad + \quad
  \parbox{10mm}{\begin{fmfchar*}(10,10)
 \fmfpen{thin}
 \fmfset{wiggly_len}{2mm}
 \fmfleft{i1} \fmfright{o1}
 \fmflabel{$k$}{i1}
 \fmflabel{$l)$}{o1}
 \fmf{wiggly}{i1,v1}
 \fmf{plain}{v1,o1}
 \fmfdot{i1}
 \fmfv{decor.shape=cross,decoration.size=2.5mm}{v1}
 \fmfdot{o1}
 \end{fmfchar*}}
 \quad \end{array} \right)
\end{equation}
where a label $k$ means the site $z_k$ and
\begin{equation}\label{91}
\fl
\quad \parbox{10mm}{\begin{fmfchar*}(10,10)
 \fmfpen{thin}
 \fmfset{wiggly_len}{2mm}
 \fmfleft{i1} \fmfright{o1}
 \fmflabel{$k$}{i1}
 \fmflabel{$l$}{o1}
 \fmf{dbl_wiggly}{i1,o1}
 \fmfdot{i1}
 \fmfdot{o1}
 \end{fmfchar*}}
 \quad  = \quad
\parbox{15mm}{\begin{fmfchar*}(10,10)
 \fmfpen{thin}
 \fmfset{wiggly_len}{2mm}
 \fmfleft{i1} \fmfright{o1}
 \fmflabel{$k$}{i1}
 \fmflabel{$l$}{o1}
 \fmf{wiggly}{i1,o1}
 \fmfdot{i1}
 \fmfdot{o1}
 \end{fmfchar*}}
   + \quad
\parbox{15mm}{\begin{fmfchar*}(15,10)
 \fmfpen{thin}
 \fmfset{wiggly_len}{2mm}
 \fmfleft{i1} \fmfright{o1}
 \fmflabel{$k$}{i1}
 \fmflabel{$l$}{o1}
 \fmf{wiggly}{i1,v1}
 \fmf{plain}{v1,v2}
 \fmf{wiggly}{v2,o1}
 \fmfdot{i1}
 \fmfdot{v1}
 \fmfdot{v2}
 \fmfdot{o1}
 \end{fmfchar*}}
  \quad + \quad
\parbox{15mm}{\begin{fmfchar*}(15,10)
 \fmfpen{thin}
 \fmfset{wiggly_len}{2mm}
 \fmfleft{i1} \fmfright{o1}
 \fmflabel{$k$}{i1}
 \fmflabel{$l$}{o1}
 \fmf{wiggly}{i1,v1}
 \fmf{plain}{v1,v2}
 \fmf{wiggly}{v2,o1}
 \fmfdot{i1}
 \fmfdot{v2}
 \fmfv{decor.shape=cross,decoration.size=2.5mm}{v1}
 \fmfdot{o1}
 \end{fmfchar*}}
 \quad + \quad
\parbox{15mm}{\begin{fmfchar*}(15,10)
 \fmfpen{thin}
 \fmfset{wiggly_len}{2mm}
 \fmfleft{i1} \fmfright{o1}
 \fmflabel{$k$}{i1}
 \fmflabel{$l$}{o1}
 \fmf{wiggly}{i1,v1}
 \fmf{plain}{v1,v2}
 \fmf{wiggly}{v2,o1}
 \fmfdot{i1}
 \fmfdot{v1}
 \fmfv{decor.shape=cross,decoration.size=2.5mm}{v2}
 \fmfdot{o1}
 \end{fmfchar*}}
\end{equation}
Here the diagrams are interpreted as in eqs. (\ref{71}), (\ref{75}),
(\ref{78}). Since the quaternion determinant is a Pfaffian and the
crossed terms factorize, it is easy to see due to $\eta^2 = 0$ for a
Grassmannian $\eta$, that only the first order terms in an expansion
of powers  of the cross contribute. We also have not to worry about
the $\delta$-functions which appear inside  the quaternion
determinant in all elements except \hspace*{2mm}
 \parbox{12mm}{\begin{fmfchar*}(10,10)
 \fmfpen{thin}
 \fmfset{wiggly_len}{2mm}
 \fmfleft{i1} \fmfright{o1}
 \fmf{plain}{i1,o1}
 \fmfdot{i1}
 \fmfdot{o1}
 \end{fmfchar*}}
, since we see from the expansion of the Pfaffian that each
$\delta$-function appears at most once at each site. The
$\delta$-functions single out special correlations of real
eigenvalues $(\delta (y_k))$ or pairs of complex conjugate
eigenvalues $(\delta^2 (z_i - \overline{z}_j))$.
\subsection{Explicit expressions}
Let us write again explicitly  the $n$-point densities as quaternion
determinants
\begin{equation}\label{92}
\fl  R_n (z_1,\ldots, z_n) = (-1)^n \mbox{ qdet } \left(
\begin{array}{cccccc}
G(1,1) & \cdots &G(1,n) & K(1,1) & \cdots & K(1,n)\\
\vdots & \ddots & \vdots& \vdots & \ddots & \vdots\\
G(n,1) & \cdots &G(n,n) & K(n,1) & \cdots & K(n,n)\\
W(1,1) & \cdots &W(1,n) & G(1,1) & \cdots & G(n,1)\\
\vdots & \ddots & \vdots& \vdots & \ddots & \vdots\\
W(n,1) & \cdots &W(n,n) & G(1,n) & \cdots & G(n,n)\\
\end{array} \right)
\end{equation}
with
\begin{eqnarray} \label{98a}
\fl
  \nonumber G(1,2) &=& \quad
\parbox{20mm}{\begin{fmfchar*}(15,10)
 \fmfpen{thin}
 \fmfset{wiggly_len}{2mm}
 \fmfleft{i1} \fmfright{o1}
 \fmflabel{$1$}{i1}
 \fmflabel{$2$}{o1}
 \fmf{plain}{i1,v1}
 \fmf{wiggly}{v1,o1}
 \fmfdot{i1}
 \fmfdot{o1}
 \fmfdot{v1}
 \end{fmfchar*}}
  + \quad
\parbox{25mm}{\begin{fmfchar*}(15,10)
 \fmfpen{thin}
 \fmfset{wiggly_len}{2mm}
 \fmfleft{i1} \fmfright{o1}
 \fmflabel{$1$}{i1}
 \fmflabel{$2$}{o1}
 \fmf{plain}{i1,v1}
 \fmf{wiggly}{v1,o1}
 \fmfv{decor.shape=cross,decoration.size=2.5mm}{v1}
 \fmfdot{i1}
 \fmfdot{o1}
 \end{fmfchar*}}
  \\
  \fl
  \nonumber K(1,2) &=& {\cal K}_N(1,2) = \quad
\parbox{25mm}{\begin{fmfchar*}(15,10)
 \fmfpen{thin}
 \fmfset{wiggly_len}{2mm}
 \fmfleft{i1} \fmfright{o1}
 \fmflabel{$1$}{i1}
 \fmflabel{$2$}{o1}
 \fmf{plain}{i1,o1}
 \fmfdot{i1}
 \fmfdot{o1}
 \end{fmfchar*}}
  \\
  \fl
  W(1,2) &=&
  \quad
\parbox{15mm}{\begin{fmfchar*}(10,10)
 \fmfpen{thin}
 \fmfset{wiggly_len}{2mm}
 \fmfleft{i1} \fmfright{o1}
 \fmflabel{$1$}{i1}
 \fmflabel{$2$}{o1}
 \fmf{wiggly}{i1,o1}
 \fmfdot{i1}
 \fmfdot{o1}
 \end{fmfchar*}}
   + \quad
\parbox{22mm}{\begin{fmfchar*}(22,10)
 \fmfpen{thin}
 \fmfset{wiggly_len}{2mm}
 \fmfleft{i1} \fmfright{o1}
 \fmflabel{$1$}{i1}
 \fmflabel{$2$}{o1}
 \fmf{wiggly}{i1,v1}
 \fmf{plain}{v1,v2}
 \fmf{wiggly}{v2,o1}
 \fmfdot{i1}
 \fmfdot{v1}
 \fmfdot{v2}
 \fmfdot{o1}
 \end{fmfchar*}}
  \quad + \quad
\parbox{22mm}{\begin{fmfchar*}(22,10)
 \fmfpen{thin}
 \fmfset{wiggly_len}{2mm}
 \fmfleft{i1} \fmfright{o1}
 \fmflabel{$1$}{i1}
 \fmflabel{$2$}{o1}
 \fmf{wiggly}{i1,v1}
 \fmf{plain}{v1,v2}
 \fmf{wiggly}{v2,o1}
 \fmfdot{i1}
 \fmfdot{v2}
 \fmfv{decor.shape=cross,decoration.size=2.5mm}{v1}
 \fmfdot{o1}
 \end{fmfchar*}}
 \quad + \quad
\parbox{22mm}{\begin{fmfchar*}(22,10)
 \fmfpen{thin}
 \fmfset{wiggly_len}{2mm}
 \fmfleft{i1} \fmfright{o1}
 \fmflabel{$1$}{i1}
 \fmflabel{$2$}{o1}
 \fmf{wiggly}{i1,v1}
 \fmf{plain}{v1,v2}
 \fmf{wiggly}{v2,o1}
 \fmfdot{i1}
 \fmfdot{v1}
 \fmfv{decor.shape=cross,decoration.size=2.5mm}{v2}
 \fmfdot{o1}
 \end{fmfchar*}}
\end{eqnarray}
The quaternion determinant of a matrix $C$ is according to eq.
(\ref{84}) up to  an overall sign equal to the Pfaffian of $J \cdot
C$. The sign is in our case easy to find since all $R_n \geqslant
0$. $G(1,2)$ may be split into two parts depending on wether the
second argument is real or complex:
\begin{equation}\label{99a}
    G(1,2) = G^C(1,2) + G^R(1,2) \delta(y_2)
\end{equation}
Then we obtain
\begin{eqnarray}
\label{100a}
 \fl \nonumber    G^C(1,2) &=& {\cal K}_N(z_1,\overline{z_2}) 2 i \mbox{ sgn } (y_2) {\rm e}^{-x_2^2 + y_2^2} \mbox{ erfc}
  (|y_2| \sqrt{2}) \\
   &=& \frac{z_1 - \overline{z_2}}{2 \sqrt{2 \pi}} {\rm e}^{z_1 \cdot \overline{z_2}} \int_{z_1 \cdot \overline{z_2}}^{\infty} du \, {\rm e}^{-u}
   \frac{u^{N-2}}{(N-2)!} \cdot 2 i \mbox{ sgn } (y_2) {\rm e}^{-x_2^2 + y_2^2} \mbox{ erfc}
  (|y_2| \sqrt{2})
\end{eqnarray}
analytic in $N$. In the second part $G^R(1,2)$ we can make a partial
integration like in section \ref{4b} which cancels the term with the
cross, which is present only for odd $N$, and obtain
\begin{eqnarray}
  \nonumber G^R(1,2) &=& -\frac{1}{\sqrt{2 \pi}} {\rm e}^{-x_2^2 + x_2 z_1} \int_{x_2 \cdot z_1}^{\infty} du \, {
  \rm e}^{-u} \frac{u^{N-2}}{(N-2)!} \\
  \label{101a}
   && -\frac{1}{\sqrt{2 \pi}} {\rm e}^{-x_2^2/2} \int_{0}^{x_2} dx \, {
  \rm e}^{-x^2/2} \frac{x^{N-2}}{(N-2)!} z_1^{N-1}
\end{eqnarray}
which is analytic in $N$ and valid for even and odd $N$.\\

\noindent  Finally we consider $W(1,2)$ which splits into $4$ parts
\begin{eqnarray}
  \nonumber W(1,2) &=& W^{CC} (1,2) + \delta(y_1) W^{RC}(1,2)  + W^{CR} (1,2) \delta( y_2) \\
   && + \delta(y_1) \delta(y_2) W^{RR} (1,2)
\end{eqnarray}
First we obtain $W^{CC} (1,2)$:
\begin{eqnarray}
 \fl  \nonumber W^{CC} (1,2) &=& - 2i \mbox{ sgn}(y_1) {\rm e}^{-x_1^2  + y_1^2} \mbox{ erfc}( |y_1| \sqrt{2})
  \left\{ \delta^2 (z_1 - \overline{z_2}) \right. \\
 \nonumber   && + {\cal K}_N (\overline{z_1}, \overline{z_2}) 2i \mbox{ sgn} (y_2) {\rm e}^{-x_2^2 + y_2^2}
   \left. \mbox{ erfc} ( |y_2| \sqrt{2}) \right\} \\
   &=& - 2i \mbox{ sgn}(y_1) {\rm e}^{-x_1^2  + y_1^2} \mbox{ erfc}( |y_1| \sqrt{2}) \left\{ \delta^2 (z_1 - \overline{z_2})
   + G^C(\overline{z_1},z_2) \right\}.
\end{eqnarray}
Again in $W^{RC}(1,2)$ and $W^{CR} (1,2)$ we can make a partial
integration to cancel the cross term and obtain
\begin{equation}\label{108a}
\fl W^{RC}(1,2) = - W^{CR} (2,1) =  2i \mbox{ sgn}(y_2) {\rm
e}^{-x_2^2 + y_2^2} \mbox{ erfc}(|y_2|\sqrt{2} )
G^R(\overline{z_2},x_1)
\end{equation}
again valid for even and odd $N$ and analytic in $N$.\\
Finally the most complicated term $W^{RR} (1,2)$ can also be reduced
to $G^R(1,2)$. One splits off a factor ${\rm e}^{-(x_1^2 +
x_2^2)/2}$ and derives a first order  differential equation in
$x_1$. Using the skew-symmetry and again a partial integration to
cancel the cross-terms one arrives at
\begin{eqnarray}
 \nonumber W^{RR}(1,2) &=& 2 \int_{x_2}^{x_1} dx \; {\rm e}^{-(x_1^2 + x^2)/2} G^R(x,x_2) - {\rm e}^{-(x_1^2 + x_2^2)/2}
  \mbox{ sgn}(x_2 - x_1) \\
 \nonumber   &=& - \int_{x_1}^{x_2} dx \; \left[ {\rm e}^{-(x_2^2 + x^2)/2} G^R(x,x_1) + {\rm e}^{-(x_1^2 + x^2)/2}
 G^R(x,x_2) \right]\\ \label{109a}
&& -  {\rm e}^{-(x_1^2 + x_2^2)/2}  \mbox{ sgn}(x_2 - x_1) \, .
\end{eqnarray}
Again this expression is valid for even and odd $N$ and analytic in
$N$. The second line is a nontrivial consequence of the
skew-symmetry of $W^{RR} (1,2)$. In all cases the analyticity in $N$
is easily seen for arguments $z_1,z_2, \ldots, z_n$ positive, but
then can be
extended.\\

\noindent At the end of this section let us write down some
correlations in the notations of this section:
\begin{eqnarray}
  R_1 (1) &=& -G(1,1) \\
  \label{107a}
  R_2 (1,2) &=&  G(1,1) G(2,2) - G(1,2) G(2,1) - W(1,2) K(2,1)
\end{eqnarray}
The last two terms yield the  connected part, i.e. the cluster
function
\begin{equation}\label{111a}
     R_2^{con} (1,2) =  - G(1,2) G(2,1) - W(1,2) K(2,1) \, .
\end{equation}
In general one draws  all possible diagrams with elements $G$, $K$,
$W$ and  the sign $(-1)^{\mbox{\small number of fermion loops}}$.
\subsection{Numerical evaluation}
\begin{figure}[t]
\begin{center}
\includegraphics[width=7cm]{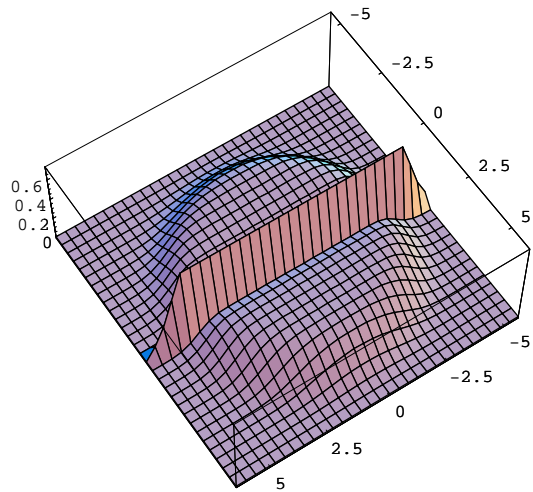}
\caption{\label{fig1} 1-point density $R_1(z_1)$ from simulation for
$N = 20$. The total integral is normalized to $N$.}
\end{center}
\end{figure}

\begin{figure}[t]
\begin{center}
\includegraphics[width=6cm]{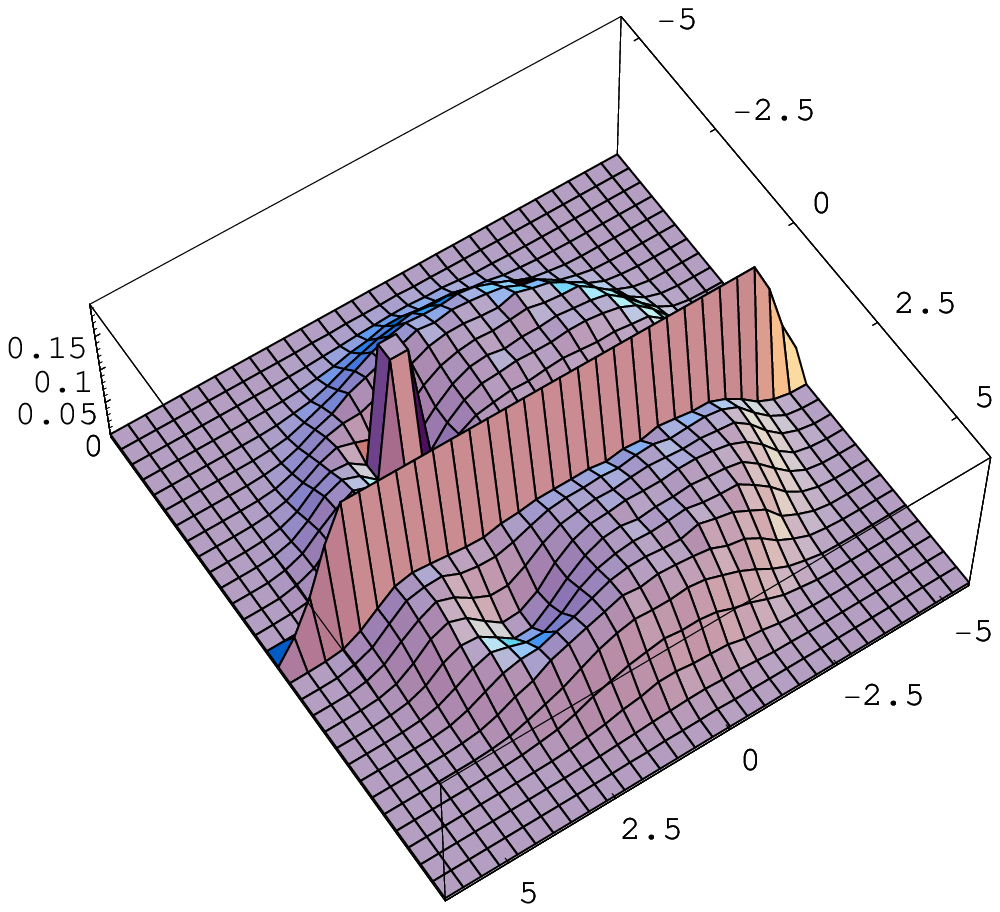}
\includegraphics[width=6cm]{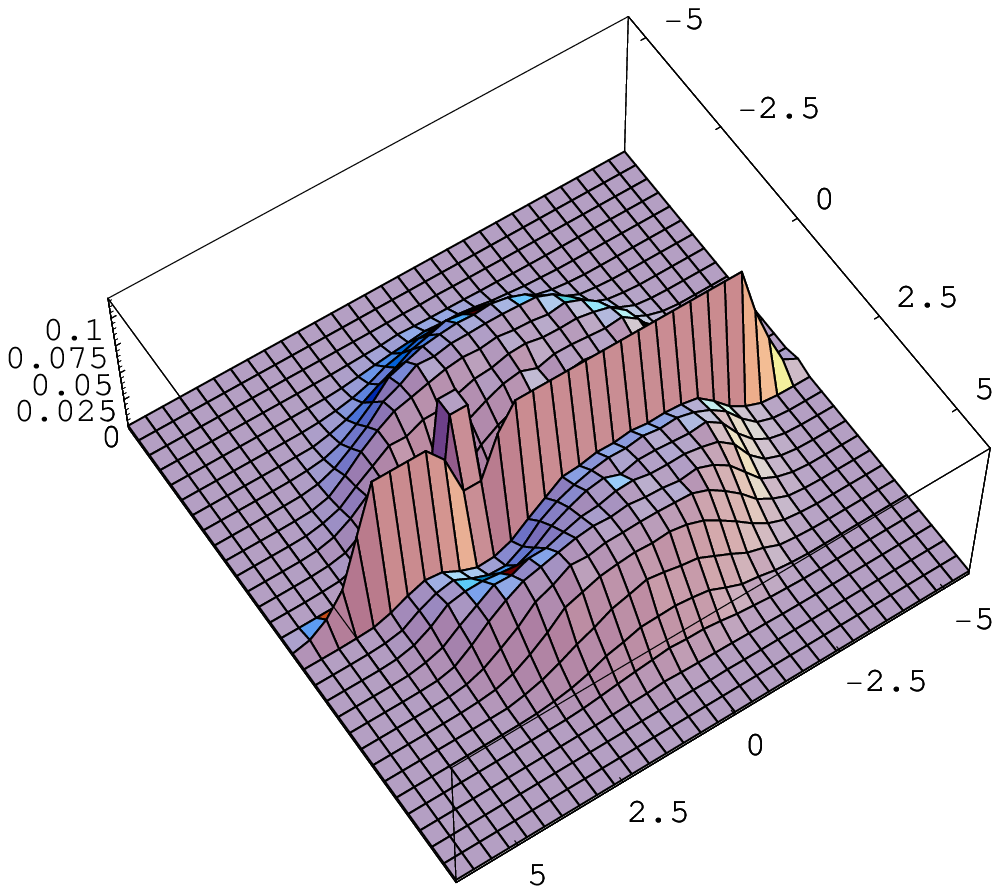}
\caption{\label{fig2} Left: 2-point density $R_2(z_1,z_2)$ as
function of $z_1$, for fixed $z_2 = 2 + 2i$ and the same simulation
data as in Fig. \ref{fig1} with $N=20$. The total integral is
normalized  to $(N-1) \cdot R_1(z_2)$. Right: The same
  for a simulation with $N=15$ and $z_2 = 2 + 0.5 i$. }
\end{center}
\end{figure}

\begin{figure}[t]
\begin{center}
\includegraphics[width=7cm]{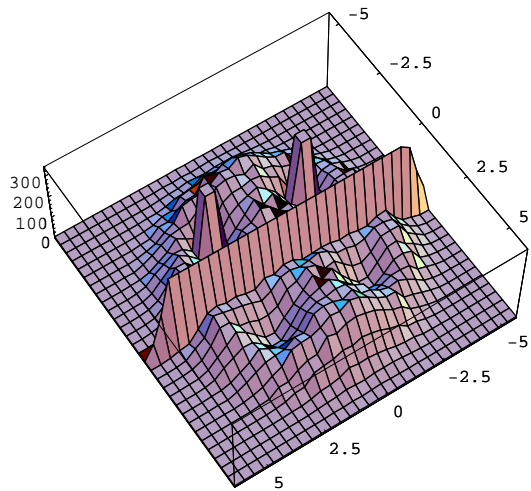}
\caption{\label{fig2a} 3-point density $R_3(z_1, z_2,z_3)$ from
simulation for $N=20$ and fixed $z_2 = 2 + 2i$, $z_3 = -2 + 2i$ as
function of $z_1$ (unnormalized). }
\end{center}
\end{figure}

\begin{figure}[t]
\begin{center}
\includegraphics[width=7cm]{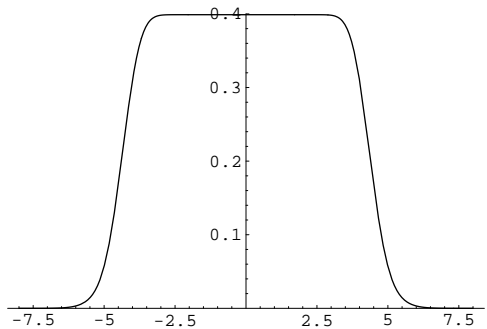}
\caption{\label{fig3} Analytical result for $R_1^R (x_1)$ for $N =
20$ as a function of $x_1$ from eq. (\ref{118a}).}
\end{center}
\end{figure}

\begin{figure}[t]
\begin{center}
\includegraphics[width=7cm]{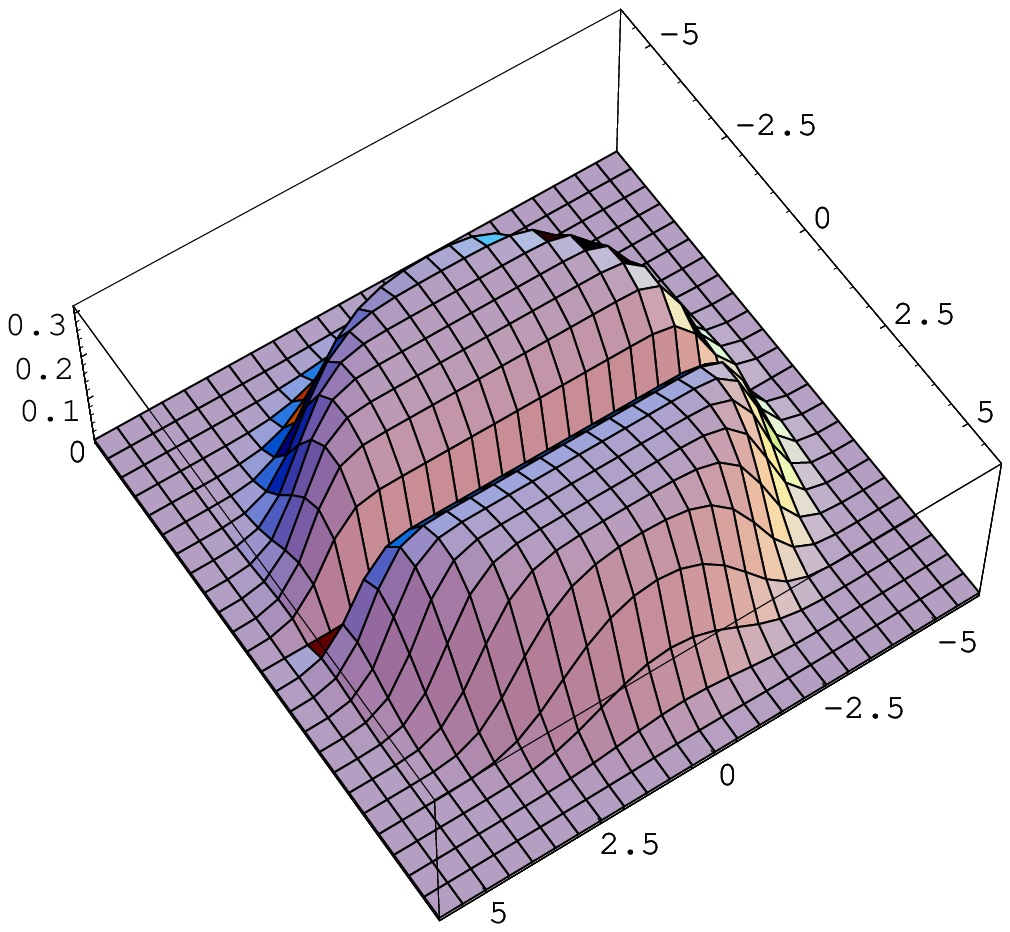}
\caption{\label{fig4} Analytical result for $R_1^C (z_1)$ for $N =
20$ as a function of $z_1$ from eq. (\ref{115}).}
\end{center}
\end{figure}

\begin{figure}[t]
\begin{center}
\includegraphics[width=6cm]{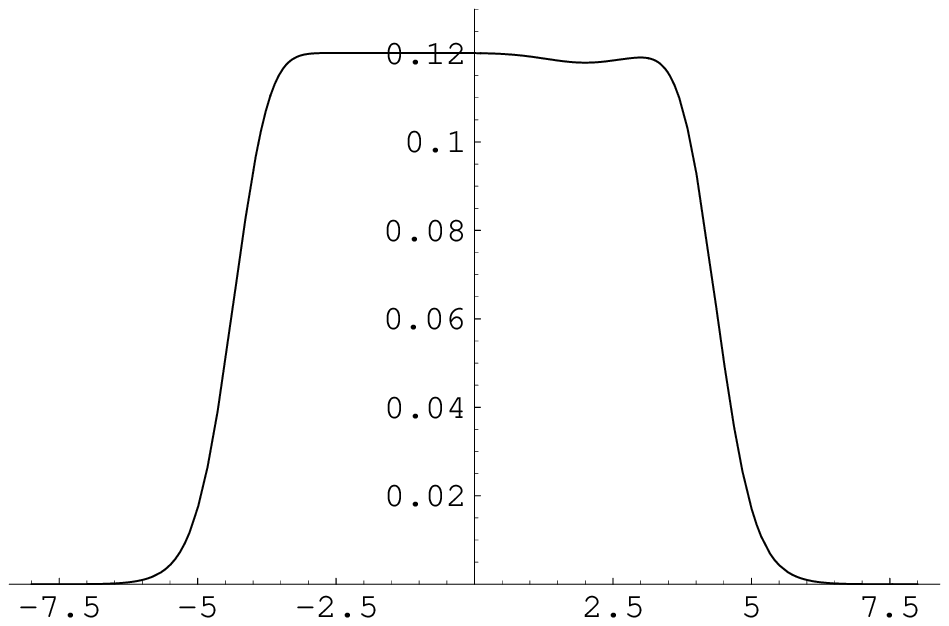}
\includegraphics[width=6cm]{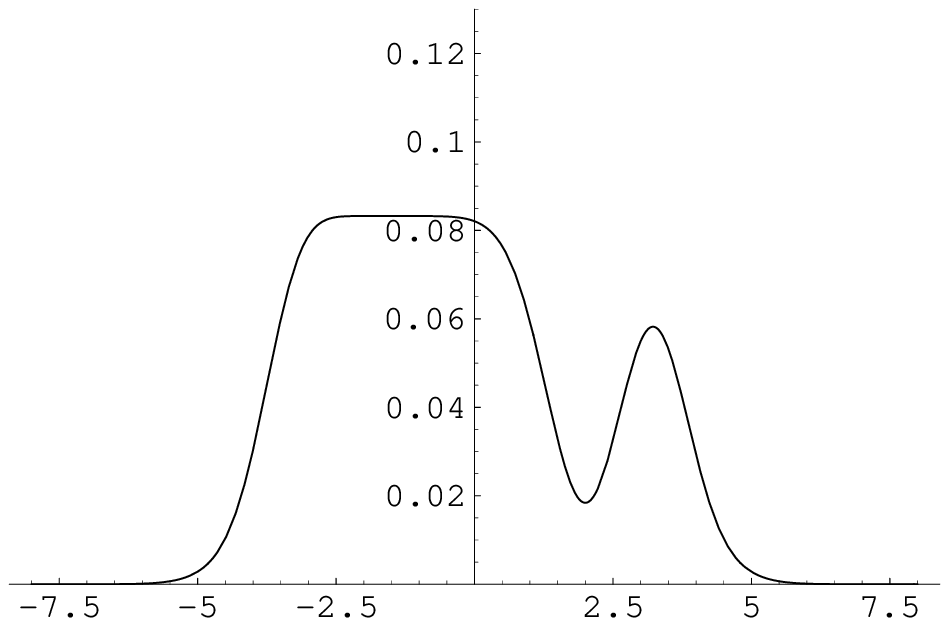}
\caption{\label{fig5} Left: Analytical  result for the correlation
$R_2^{RC} (x_1,z_2)$ for $N = 20$ and fixed $z_2 = 2 + 2 i$ as a
function of the real $x_1$ from eq. (\ref{119a}). Right: The same
for $N = 15$ and $z_2 = 2 + 0.5 i$. }
\end{center}
\end{figure}

\begin{figure}[t]
\begin{center}
\includegraphics[width=6cm]{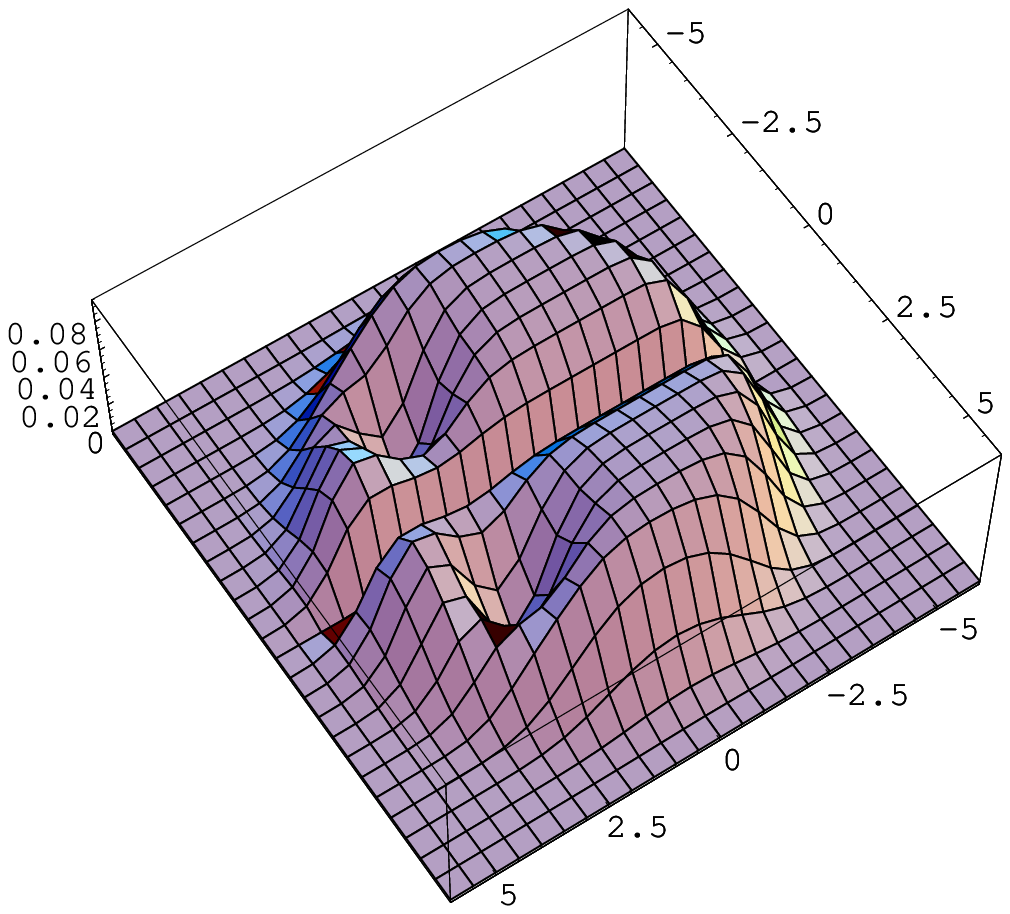}
\includegraphics[width=6cm]{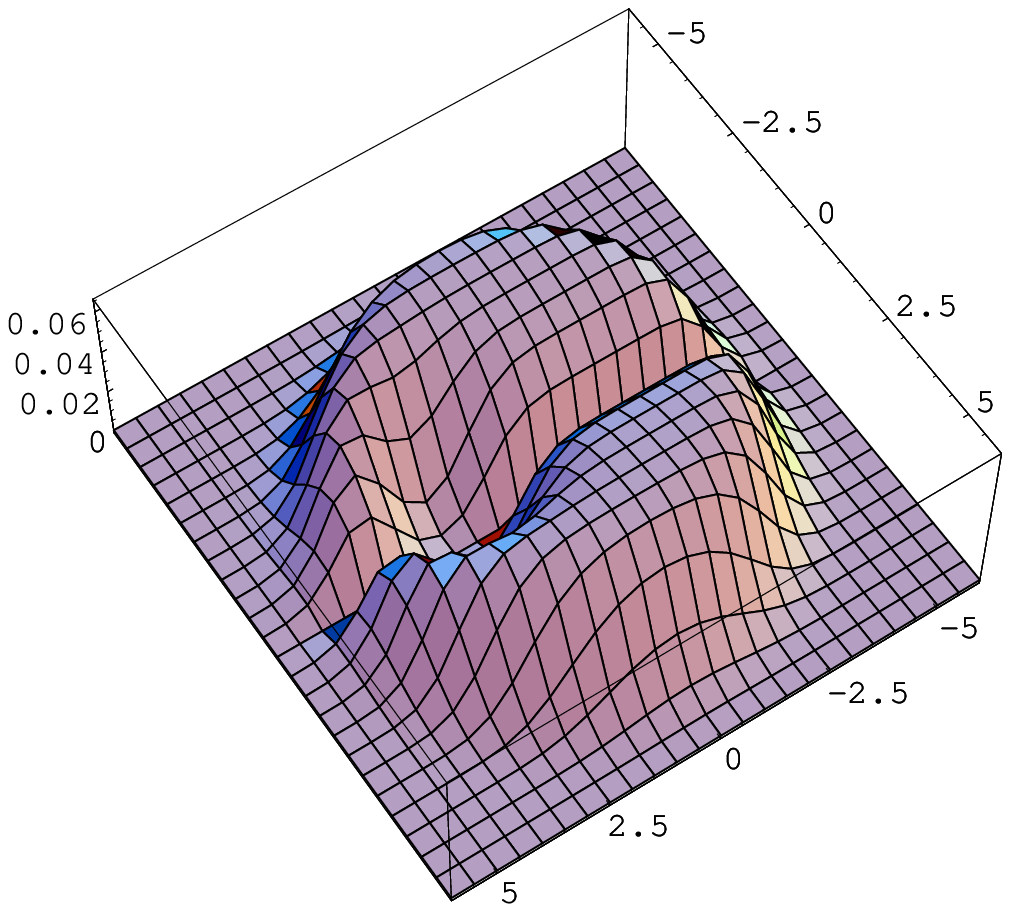}
\caption{\label{fig6} Left: Analytical  result for the correlation
$R_2^{CC} (z_1,z_2)$ for $N = 20$ and fixed $z_2 = 2 + 2 i$ as a
function of  $z_1$ from eq. (\ref{119a}). Right: The same for $N =
15$ and $z_2 = 2 + 0.5 i$.  }
\end{center}
\end{figure}

\begin{figure}[t]
\begin{center}
\includegraphics[width=7cm]{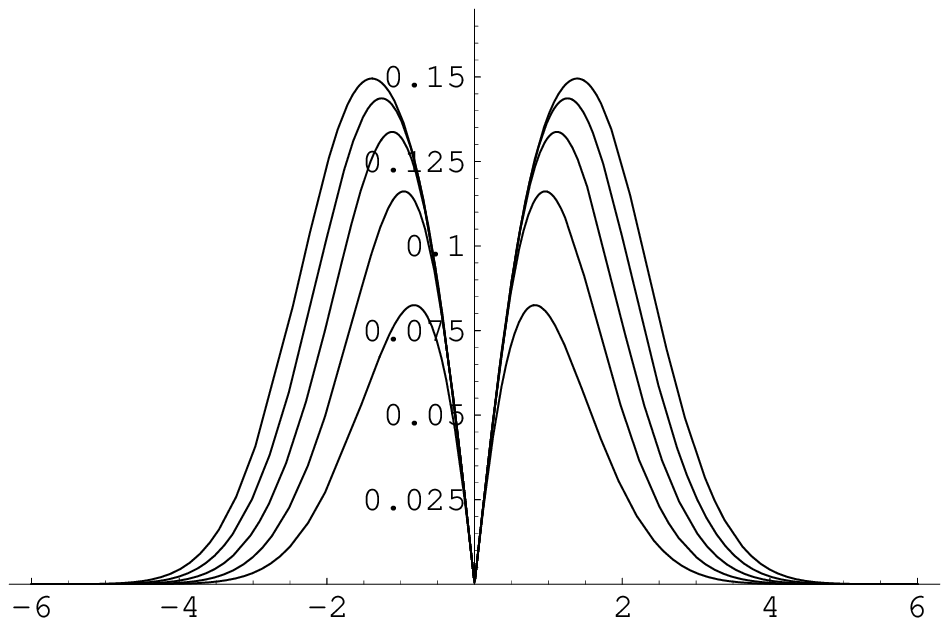}
\caption{\label{fig7} Analytical  result for the correlation
$R_2^{RR} (x,0)$ for $N = 2,3,4,5,6$ from eq. (\ref{119}). }
\end{center}
\end{figure}

Let us introduce the function $\phi (z)$:
\begin{equation}\label{107}
    \phi (z) = - 2i \mbox{ sgn}(y) \, {\rm e}^{-x^2 + y^2} \mbox{ erfc} (| y | \sqrt{2})
    = - \phi (\overline{z})
\end{equation}
and the incomplete Gamma function $\gamma^{*} (n,x)$:

\begin{equation}\label{109}
    \Gamma(n) \, x^n \gamma^{*} (n,x)  = \int_{0}^{x} du \, {\rm e}^{-u} u^{n-1} \, .
\end{equation}
$\gamma^{*} (n,x)$ is an analytic function of both arguments. It has
the power expansion
\begin{equation}\label{108}
\gamma^{*} (n,x) = {\rm e}^{-x} \sum_{m=0}^{\infty}
\frac{x^m}{\Gamma (n + m +1)} \, .
\end{equation}
 Then we obtain as basic functions from eqs.
(\ref{54a}),(\ref{54}), (\ref{100a}), (\ref{101a})
\begin{equation}\label{110}
    K(1,2) = {\cal K}_N (z_1, z_2) = \frac{z_1 - z_2}{2 \sqrt{2\pi}} {\rm e}^{z_1 z_2} \bigl(
    1 - (z_1 z_2)^{N-1} \gamma^{*} (N-1, z_1 z_2) \bigr)
\end{equation}
and
\begin{equation}\label{111}
    G^C(1,2) = G^C(z_1,z_2) = - {\cal K}_N (z_1, \overline{z_2}) \phi (z_2)
\end{equation}
and
\begin{eqnarray}
\label{112}
  \nonumber  G^R(1,2) &=& G^R(z_1,x_2) = - \frac{1}{ \sqrt{2\pi}} {\rm e}^{- x_2^2 + x_2 z_1} \bigl(
    1 - (z_1 x_2)^{N-1} \gamma^{*} (N-1, z_1 x_2) \bigr) \\
  && - \frac{{\rm e}^{- x_2^2/2} (z_1 x_2)^{N-1} }{2^{N - 1/2} \Gamma (N/2)}  \gamma^{*} \Bigl(\frac{N-1}{2}, \frac{x_2^2}{2}\Bigr)
   \, .
\end{eqnarray}
From this we obtain
\begin{eqnarray}
  \nonumber W^{CC} (1,2)  &=&  \phi (z_1) ( \delta^2(z_1 - \overline{z_2}) + G^C(\overline{z_1},z_2) )\\
  &=& \phi (z_1) \delta^2 (z_1 - \overline{z_2} ) + \widetilde{W}^{CC} (1,2)
\end{eqnarray}
The first term leads to a self-correlation of a complex conjugate
pair. Then
\begin{equation}\label{114}
    W^{RC} (1,2) = - W^{CR} (2,1) = - \phi(z_2) G^R(\overline{z_2}, x_1)
\end{equation}
and $W^{RR} (1,2)$ is given by eq. (\ref{109a}). These formulae
immediately yield  all $n$-point densities:
\begin{eqnarray}
\label{115}
  R_1^{C} (1) = - G^{C} (1,1) &=& {\cal K}_N (z_1,\overline{z_1}) \phi (z_1) \\
  \nonumber  R_1^{R} (1) = - G^{R} (1,1) &=& \frac{1}{\sqrt{2 \pi}}  \bigl(
    1 - x_1^{2(N-1)} \gamma^{*} (N-1, x_1^2 ) \bigr) +\\
    \label{118a}
&&  + \frac{{\rm e}^{-x_1^2/2} x_1^{2 (N-1)}}{2^{N - 1/2} \Gamma
(N/2)} \gamma^{*} \Bigl( \frac{N-1}{2}, \frac{x_1^2}{2}\Bigr)
\end{eqnarray}
and also $R_2 (1,2)$ splits into 5 parts:
\begin{eqnarray}
\fl   \nonumber R_2(1,2) &=& R_1^C (1) \delta^2 (z_1 -
\overline{z_2}) + R_2^{CC} (1,2) + R_2^{RC}(1,2) \delta (y_1)
  + R_2^{CR} (1,2) \delta(y_2)\\
  && + R_2^{RR} (1,2) \delta(y_1) \delta(y_2) \, .
\end{eqnarray}
All the terms follow from eq. (\ref{107a}). The first term
corresponds to the correlation of a complex eigenvalue with its
complex conjugate. The other smooth terms correspond to correlation
complex-complex, real-complex, complex-real, real-real. For
completeness let us write down these terms:
\begin{eqnarray}
\label{119a}
\fl \nonumber R_2^{CC} (1,2) = G^C (1,1) G^C(2,2) - G^C (1,2) G^C(2,1) - \widetilde{W}^{CC} (1,2) K(2,1) \\
\fl \nonumber R_2^{RC} (1,2) = G^R (1,1) G^C(2,2) - G^C (1,2) G^R(2,1) - W^{RC} (1,2) K(2,1) \\
\fl \nonumber R_2^{CR} (1,2) = G^C (1,1) G^R(2,2) - G^C (2,1) G^R(1,2) - W^{CR} (1,2) K(2,1) \\
 \fl R_2^{RR} (1,2) = G^R (1,1) G^R(2,2) - G^R (1,2) G^R(2,1)- W^{RR}
(1,2) K(2,1)
\end{eqnarray}
In the following we do some numerical simulations and compare them
with numerical evaluations of the above formulae. We draw randomly
matrices $J_{ij}$ from the Gaussian ensemble (\ref{1}) and plot at
first a histogram for the eigenvalues in the complex plane, which
yields $R_1(z_1)$ (Fig. \ref{fig1}).
We see that the eigenvalues lie in a circle with radius of order
$\sqrt{N}$ and that a finite  fraction lies strictly on the real
axis repelling the remaining pairs of complex conjugate eigenvalues
from the real axis. Then we take the same set of eigenvalues and
choose only that subset with one eigenvalue close to a fixed value
$z_2$. Plotting a histogram of this set we obtain $R_2(z_1,z_2)$
(Fig. \ref{fig2}).
 We see that the complex eigenvalue $z_2$ repells all
the other  with a cubic law in distance and that there is again a
finite fraction of eigenvalues on the real axis corresponding to
complex-real correlation. Furthermore one finds a $\delta$-peak at
the complex conjugate site $\overline{z_2}$ and also repulsion from
that point. We can even take the same data, fix two eigenvalues
$z_2$ and $z_3$ and plot $R_3  (z_1, z_2 ,z_3)$ as
function of $z_1$ (Fig. \ref{fig2a}).\\
\noindent Using the above formulae we can calculate $R_1^C (1)$,
$R_1^R (1)$, $R_2^{CC} (1,2)$, $R_2^{RC}(1,2)$, $R_2^{RR} (1,2)$
etc. exactly. Below we plot the functions $R^R_1(x_1)$, $R_1^C
(z_1)$, $R_2^{RC} (x_1,z_2)$ and $R_2^{CC} (z_1,z_2)$ for the same
fixed $z_2$ as in the simulation (Fig. \ref{fig3}-\ref{fig6}).

A further plot (Fig. \ref{fig7}) shows the correlation $R_2^{RR}
(x,0)$ for $N= 2,3,4,5,6$, for which we have the analytical formula
from eqs. (\ref{119a}, \ref{112}, \ref{109a}, \ref{110}):
\begin{equation}\label{119}
    R_2^{RR} (x,0)  = \frac{1}{\sqrt{2\pi}} R_1^R(x) - \frac{1}{2\pi} {\rm e}^{-x^2}
      + \frac{|x|}{2\sqrt{2\pi}} {\rm e}^{-x^2/2} \mbox{ erfc}
    (\frac{|x|}{\sqrt{2}}) \, .
\end{equation}
The only $N$-dependence is in the first term  $R_1^R (x)$, which  is
almost constant in $x$ for sufficiently large $N$. For small $|x|$
we have linear level repulsion. The connected correlation at the
special point is independent of $N$ and decays with ${\rm
e}^{-x^2}$.

\section{Conclusions}
In this paper we have derived, starting from  the Gaussian ensemble,
closed analytical expressions for all correlation functions, i.e.
$n$-point densities, of eigenvalues of the real Ginibre ensemble of
real asymmetric matrices, which is invariant under real orthogonal
transformations. These $n$-point densities are not absolutely
continuous. They contain contributions which are concentrated by
$\delta$-functions on the real axis and also point-measures for
pairs of complex conjugate eigenvalues which are always present,
because the eigenvalues of a real asymmetric  matrix are either real
or pairwise complex conjugate. The $n$-point densities are written
as quaternion determinants of certain $2n \times 2n$ selfdual
matrices, or as Pfaffians of corresponding skew-symmetric matrices.
The Pfaffians can be derived from a zero-dimensional fermion field
theory, similar in structure as the matrix Green functions in the
Nambu space for superconductivity. All the $n$-point densities are
expressible in terms of a skew-symmetric measure ${\cal F}
(z_1,z_2)$ containing one part concentrated on the complex conjugate
pairs $z_1 = \overline{z_2}$ and one part on the real eigenvalues
$z_1 = \overline{z_1}$, $z_2 = \overline{z_2}$. For the
$N$-dimensional matrix $J_{ij}$ one constructs from ${\cal F}
(z_1,z_2)$ a skew-symmetric kernel ${\cal K}_N (z_1,z_2)$ which
together with ${\cal F} (z_1,z_2)$ yields the building  blocks
for the correlations.\\
 To calculate ${\cal K}_N (z_1,z_2)$ one has to invert an $N \times N$ antisymmetric matrix
 $A_{kl}$ related to ${\cal F} (z_1,z_2)$, which looks complicated, which however turns out
 to yield a very simple tridiagonal structure for $A^{-1}_{kl}$. To find this it is enough
 to compare the Edelman result for the complex 1-point density with the general form of the 1-point density. This is
 sufficient to obtain all correlations in the case of even dimension $N$. In the case of odd dimension $N$ one
 has to increase the  dimension artificially by 1 and has to invert instead the $(N+1) \times (N+1)$ dimensional
 skew-symmetric matrix $B_{nm}$. Again a simple argument using the result of Edelman, Kostlan and Shub
 for the real 1-point density reveals that $B^{-1}_{nm}$ has again a simple tridiagonal (but slightly more complicated)
 structure. In this paper we do not discuss detailed asymptotics for large $N$, which has partly been discussed elsewhere
 \cite{Sommers2007,Forrester07,Borodin08} and which follows essentially from the asymptotic kernel   ${\cal K}_N (z_1,z_2)\backsimeq (z_1 - z_2)
  {\rm e}^{+ z_1 z_2}/2\sqrt{2 \pi}$. However we show that the different formulae of the correlation functions for even and odd $N$
  can be combined  always in one formula which depends on continuous     and even analytic functions of $N$.\\

  \noindent Finally we have presented some numerical simulations which make clear the complicated
  structure of the correlations,
  for example the 2-point correlation. There is a smooth background  of complex-complex correlations, then a point measure for a
  complex conjugate pair, a part with one $\delta$-function concentrated on the real axis for complex-real
  correlations and a part corresponding to real-real correlations doubly concentrated on  the real axis. In
  comparison we have shown also some numerical evaluations of the
   analytical formulae.\\

   \noindent For the future there remain  to be discussed more detailed asymptotics and more sophisticated functions like level spacings and
   distributions of extremal eigenvalues.

\ack This project has been supported by the
Sonderforschungsbereich/Transregio 12 of the Deutsche
Forschungsgemeinschaft.

\end{fmffile}

\section*{References}
\begin{thebibliography}{10}
\bibitem{Sommers2007} H.-J. Sommers,   J. Phys. A {\bf 40}, 671  (2007)
\bibitem{Ginibre65} J.Ginibre, J. Math. Phys. {\bf 6}, 440 (1965)
\bibitem{Lehmann91} N.Lehmann and H.-J. Sommers, Phys. Rev. Lett. {\bf 67}, 941 (1991)
\bibitem{Edelman97} A.Edelman, J. Multivariate Anal. {\bf 60}, 203 (1997)
\bibitem{EKS94} A.Edelman, E.Kostlan and M.Shub, J. Amer. Math. Soc. {\bf 7}, 247 (1994)
\bibitem{Akemann07} G. Akemann and E.Kanzieper, J. Stat. Phys. {\bf 129}, 1159 (2007)
\bibitem{Sinclair06} C.D.Sinclair, arXiv: math-ph/0605006 (2006)
\bibitem{Forrester07} P.J.Forrester and T. Nagao, Phys. Rev. Lett. {\bf 99}  050603 (2007)
\bibitem{Borodin08} A.Borodin, C.D.Sinclair, arXiv: math-ph/0706.2670v2
\bibitem{May72} R.M.May, Nature {\bf 298 } 413 (1972)
\bibitem{Sompolinsky88} H.Sompolinsky, A.Crisanti and H.-J.Sommers, Phys. Rev. Lett. {\bf 61} 259 (1988)
\bibitem{Efetov97} K.B.Efetov, Phys. Rev. Lett. {\bf 79} 491 (1997)
\bibitem{KWAPIEN06} J.Kwapien, S. Drozdz, A.Z. Gorski and  F. Oswiecimka, Acta  Phys. Pol. {\bf B37}, 3039 (2006)
\bibitem{Bruzda08} W.Bruzda, V.Cappelini, H.-J.Sommers, K.\.Zyczkowski,  arXiv: quant-ph/0804.2361
\end{thebibliography}
\end{document}